# Cracking the Code: Enhancing Development finance understanding with artificial intelligence

Pierre Beaucoral[1]

Université Clermont Auvergne, CNRS, IRD, CERDI, F-63000 Clermont-Ferrand, France

2025-01-21

## Abstract

Analyzing development projects is crucial for understanding donors' aid strategies, recipients' priorities, and to assess development finance capacity to adress development issues by on-the-ground actions. In this area, the Organisation for Economic Co-operation and Development's (OECD) Creditor Reporting System (CRS) dataset is a reference data source. This dataset provides a vast collection of project narratives from various sectors (approximately 5 million projects). While the OECD CRS provides a rich source of information on development strategies, it falls short in informing project purposes due to its reporting process based on donors' self-declared main objectives and pre-defined industrial sectors. This research employs a novel approach that combines Machine Learning (ML) techniques, specifically Natural Language Processing (NLP), an innovative Python topic modeling technique called BERTopic, to categorise (cluster) and label development projects based on their narrative descriptions. By revealing existing yet hidden topics of development finance, this application of artificial intelligence enables a better understanding of donor priorities and overall development funding and provides methods to analyse public and private projects narratives.

## I. Introduction

As development projects become increasingly complex and diverse, and public opinion regarding aid in both donor and recipient countries becomes more critical, there is a growing recognition of the need for advanced methodologies to analyse and categorize them, ensuring transparency and accountability of those projects (Honig and Weaver 2019). Traditional methods of analysis frequently rely on predetermined variables, simplified categorisations, or fixed labelling systems. Thus, they may struggle to capture the intricacies embedded in project narrative descriptions, paradigm shifts, or new schemes that emerge over time. Especially if they are not detected and implemented in advance. This work aims to address the aforementioned limitation by introducing a new and easily replicable framework. Machine Learning (ML), specifically Natural Language Processing (NLP), is used to extract significant patterns directly from project narratives in the OECD CRS dataset, a commonly used data source in development finance studies (available at this link: https://stats.oecd.org/viewhtml.aspx?datasetcode=TABLE5&lang=en). This dataset serves as a reference for studying development finance and economics. It includes information on (i) public aid (Alesina and Dollar 2000; Berthélemy et al. 2006; and Clist 2011), (ii) climate Aid (Halimanjaya 2015; and Noel 2023), (iii) Climate adaptation Aid (Weiler, Klöck, and Dornan 2018). These are issues of growing concern in our societies, explaining why there is an increasing demand for more clarity and transparency in their handling by public authorities.

[1] This work was supported by the Agence Nationale de la Recherche of the French government through the program "France 2030" (grant number ANR-16-IDEX-0001).

The literature surrounding the OECD CRS dataset trying to examine global development finance is mainly concentrated on thematic analysis using dummy variables, such as Rio markers or sector codes, to categorise projects (Lee and Lim 2014; Betzold 2016; and Han and Cheng 2023). Furthermore, sectoral categorisation may be appropriate for industrial practices, but may give a distorted view of aid that is not fully comparable to trade and private sector operations. As an example, since 1998, the Development Assistance Committee (DAC) of the Organisation for Economic Co-operation and Development (OECD) has established the Rio markers system (*Creditor Reporting System on Aid Activities* 2002). This system is composed of policy markers that monitor and report on development finance flows targeting the themes of the Rio Conventions. The four markers are (i) biodiversity, (ii) desertification, (iii) climate change mitigation, and (iv) adaptation. To target one of these markers in a project, implementing agencies must declare that one of these themes is a significant or principal objective of the project. The majority of studies and reports on these topics have traditionally relied on this quite straightforward classification when using this reference dataset (Betzold and Weiler 2018; Halimanjaya 2014; and Betzold 2016). Although these studies, including the ones cited earlier, have provided valuable insights into specific themes, such as poverty reduction or climate mitigation, they often overlook the wealth of information encapsulated within the narrative descriptions of each project, another set of variables available in the OECD CRS dataset. This overlook may be a missed opportunity to take into account the complete available information on development projects and development financial flows. As a consequence, it limits the depth of understanding and may fail to take in critical aspects of project dynamics. This lack of information about the implementation of development projects and their objectives could also be an opportunity to raise public awareness about the use of public spending at a time of budgetary constraints in developed economies. The contribution of this work is threefold. First, it is aiming to fill this aforementioned gap using the complete information embedded in donors' declarations thanks to project descriptions. Indeed, categorising project narrative descriptions at a deeper level may raise questions about development projects and the study of multiple development sub-topics, such as financing crops or responding to natural disasters. Second, this work aims to facilitate research and policymaking to investigate specific subjects related to development finance, enhancing our understanding of each type of financial development flow. Third, it takes part in an expanding literature on the use of machine learning to evaluate the financing of sustainable development goals (e.g. Burgess, Custer, and Custer 2023; and Toetzke, Banholzer, and Feuerriegel 2022) and machine learning-based approach to perform topic modeling on a text corpus.

Initially, as a development economist, the idea was to replicate the process outlined in Toetzke, Banholzer, and Feuerriegel (2022) to conduct further studies. However, significant advancements in machine learning and natural language processing methods led to believe that one could enhance the original method to provide a more efficient and above all more easily replicable approach for clustering development aid projects. The reproducibility of this new proposed method is essential to this work to provide a more insightful pattern regarding development finance that might be used for further research and policy implications. To provide the most comprehensive view of development finance, it is also aimed to expand their initial work to encompass the entire OECD CRS dataset as they focus on a restricted part of the dataset. Through this novel approach, this study seeks to demonstrate that the richness of project narratives can serve as a valuable resource for a deeper exploration of development initiatives. This shift from predefined variables to narrative-driven analysis has the potential to uncover hidden patterns in and connections between development projects. Therefore, this study endeavors to provide researchers with a more comprehensive view of the multifaceted nature of the development finance architecture. Additionally, this methodology aims to address possible bias across donors resulting from their different declaration habits or instructions when faced with the same repository. It is hoped this work can serve as a tool for studying multi-scalar and specific development topics and projects, from beekeeping and honey production, to large-scale nature conservation and biodiversity projects. Another objective of this work is to enable researchers to reproduce specific categories of development financial flows or to break down financial flows from different perspectives in greater details. To achieve this, all processes and results are freely available online at the following link: https://github.com/PierreBeaucoral/ML-clustering-of-development-activities.

Furthermore, in the rapidly growing field of artificial intelligence, particularly in machine learning and natural language processing, the latest methods are utilized to achieve the highest possible level of accuracy and to illustrate the potential contributions of these emerging technologies to economic academic research. As we will delve into the details of the methodology, this research aims not only to contribute to the field of development studies but also to pave the way for a more sophisticated and nuanced analysis of large-scale datasets across various domains (e.g. social media data like tweets analysis or bibliographic analysis for meta-analysis), breaking the limitations of traditional thematic categorizations, illustrating a range of possibilities for economics and social science research.

## II.    Methodology and motivations

This work builds upon the research conducted by Toetzke, Banholzer, and Feuerriegel (2022), which aimed to be replicated. However, due to significant advancements in the ML NLP field and technologies (Johri et al. 2021), it was deemed more effective to enhance the original concept. The objective of this previous work was to use NLP techniques to create accurate clusters of aid activities and label their topics based on the narratives declared in the dataset. The authors were able to classify projects into almost 180 categories, going beyond sector or objective-based classifications, such as Rio markers. In the replication process, the classification appears to be unconvincing since relying on metric (silhouette score) to assess the optimal number of clusters. However, the highest silhouette score obtained for a fixed number of clusters seemed to be poor (around 0). Yet, in this work, the use of an alternative clustering algorithm, which will be elucidated in the subsequent sections, enabled the identification of over 400 distinct and analysable thematic clusters, allowing a more refined analysis of development finance. The clustering process also seems to benefit from improvements thanks to theses changes, reflected by the different metrics used to assess for the clustering quality. This work entails several steps that are common in natural language processing (NLP), including data collection, cleaning, and preprocessing, before employing machine learning (ML) techniques.

### A.    Data Collection

The study relies on the OECD CRS dataset, a comprehensive collection of development projects, including data about the donor country, implementing agency, channel of delivery, recipient country, amount committed and disbursed yearly on the project and project narratives. The dataset spans various sectors, providing a representative sample of public development finance for analysis. The sample comprises projects declared from 1973 to 2022, totalling approximately 5 million observations. The OECD data portal provides free access to the raw yearly datasets. The OECD gathers, computes, and updates information on development projects each year based on declarations from donor and recipient entities. This study will focus on three main variables: the project title, short description, and long description, which will be concatenated in order to obtain one "raw text" variable.

## B.  Original method from Toetzke, Banholzer, and Feuerriegel (2022)

It is found to be necessary to explain the original methodology which inspired this study, in order to get a better understanding of the changes implemented and their effects on the outputs.

### 1.  The idea

The primary objective is to group projects into homogeneous clusters using project descriptions. To achieve this, it is necessary to code each project description, identify themes, and classify project descriptions according to these themes. Thus, in terms of the first main output, our work does not necessarily differ from that of Toetzke, Banholzer, and Feuerriegel (2022).

### 2.  The original model from Toetzke, Banholzer, and Feuerriegel (2022)

In their work, they used Doc2Vec, a distributed Bag of Words version of Paragraph Vector model (Le and Mikolov 2014). The goal of the algorithm is to create a numeric representation of a document, regardless of its length. The vectors that represent similar documents will be closed points in the created vector space. In the end, this algorithm allows to represent all the documents $D$ in a high-dimensional vector space. The authors' original method trains the model three times: once on a large corpus of Wikipedia articles redacted in English (approximately representing 20 gigabytes of compressed data), once on project descriptions, and once on both simultaneously.

### 3.  Their clustering process

In their original article, they use K Means Clustering. K-Means Clustering is a method of vector quantization from signal processing. It aims to partition $n$ observations into $k$ clusters, with each observation belonging to the cluster with the nearest mean (cluster centers or cluster centroids) serving as a prototype of the cluster. It is beneficial to note that this method necessitates the manual specification of k, which can prove to be challenging in identifying the optimal number of clusters. Once done, the data space is partitioned into Voronoi cells. A Voronoi cell is a part of a voronoi diagramm where each point from this cell is closer from a given point belonging to the cell (here those given points are our centroïds) than another given point from other Voronoi cells. The within-cluster variances (squared Euclidean distances) are minimized by k-means clustering. However, it does not optimize regular Euclidean distances. The mean optimizes squared errors, whereas only the geometric median minimizes Euclidean distances.

## 4.    The output

Using this unsupervised method, the authors have identified 173 activity clusters, which is an improvement compared to conventional aid activity classifications as it allows a freer distribution of projects within categories that are not defined *a priori.* Therefore, it is necessary to determine if the quality of the clustering is sufficient enough for the output to be used. In this kind of set-up, it may represent a huge challenge as clusters are created *ex nihilo,* meaning we do not possess any counter factual to compare the clustering with. To assess the quality of the clustering process, it is then needed to get a testable definition of a good clustering. Another challenge with K-means clustering is that it forces each observation (here, project description) to be in a cluster. However, some generic projects with non-specific descriptions may be described as "noise projects" and may not help to get an idea of donors' concerns when providing aid.

## C.    What is a good clustering?

A definition of a good clustering could be a situation where (i) each object is assigned to only one cluster, (ii) each cluster is clearly defined and does not overlap with others, (iii) there are high (low) intra (inter)-class similarity within clustered objects. The first part of the definition is insured by the use of the method of K-means. Indeed, K-mean clustering forces each processed document into only one existing cluster. However, in k-means, the desired number of cluster is fixed by human and may not be optimal. In this case, it might create overlap between clusters and create sub-optimal intra and inter similarity within clusters and documents (project descriptions).

### 1.    Assessing the quality of clustering process

How can we assess the quality of clustering? As an unsupervised machine learning technique, we cannot compare our results to the 'ground truth' as it does not exist. Therefore, there are two remaining solutions: compare machine learning results with human classification (as suggested by Toetzke, Banholzer, and Feuerriegel 2022) or conduct statistical tests to ensure the three main characteristics of a good clustering process mentioned above. In the second case, three metrics may be mentioned: Silhouette score, Davies-Bouldin and Calinski-Harabasz Index

*a)* **Silhouette score**

The silhouette score (Rousseeuw 1987) is one of the most commonly used metric for assessing clustering quality. The Silhouette Coefficient is defined for each sample and is composed of two scores:

- $a$: The mean distance between a sample and all other points in the same class:
  $$a(i) = \frac{1}{|I_k|-1}\sum_{j \in I_k, j \neq i} d\left(x^i, x^j\right)$$

- $b$: The mean distance between a sample and all other points in the *next nearest cluster*:
  $$b(i) = \min_{k' \neq k} \frac{1}{|I_{k'}|}\sum_{i' \in I_{k'}} d\left(x^i, x^{i'}\right)$$

Where $d\left(x^i, x^{i'}\right)$ represents the dissimilarity between our two individuals and $I_k = i \in [\,[1, N]\,]/C(i) = k$ represents all the points belonging to a group $k$ issued from a cluster $C(i)$. The Silhouette Coefficient $s$ for a single sample is then given as:

$$s(i) = \frac{b(i) - a(i)}{max\left(a(i), b(i)\right)}$$

The Silhouette Coefficient for a set of samples is given as the mean of the Silhouette Coefficient for each sample:

$$S = \frac{1}{K}\sum_{k=1}^{K}\frac{1}{|I_k|}\sum_{i \in I_k} s\left(i\right)$$

The value of this score is defined between $-1$ and $1$. It is generally considered as good if the score is above $0,5$.

*b)* **Davies-Bouldin Index**

The Davies-Bouldin index (Davies and Bouldin 1979) calculates the average "similarity" between clusters. To do so, it measures the distance between clusters compared to their size. Practically, it is an average of the similarity $R_{ij}$ of each cluster $C_i$ and its most similar one $C_j$ :

The index is defined as:

$$DB = \frac{1}{k}\sum_{i=1}^{k} \max_{i \neq j} R_{ij},$$

with $R_{ij} = \frac{s_i + s_j}{d_{ij}}$, $s_i$ representing the cluster diameter, measured as the average distance between each point of the cluster $i$ and the centroid of that same cluster, and $d_{ij}$ representing the distance between clusters' $i$ and $j$ centroids.

The closer to zero this index is, the better the partition.

### c) *Calinski-Harabasz Index*

Also known as the Variance Ration Criterion (Calinski and Harabasz 1974) is the ratio of the sum of between-clusters dispersion and of within-cluster dispersion for all clusters (where dispersion is defined as the sum of distances squared) for a set of data $E$ of size $n_E$ clustered in $k$ clusters:

$$s = \frac{tr(B_k)}{tr(W_k)} \times \frac{n_E - k}{k - 1}$$

Where $tr(B_k)$ and $tr(W_k)$ are respectively the trace of the between and within group dispersion matrix as:

$$W_k = \sum_{q=1}^{k} \sum_{x \in C_q} \left(x - c_q\right) \left(x - c_q\right)^T$$

$$B_k = \sum_{q=1}^{k} n_q \left(c_q - c_E\right)\left(c_q - c_E\right)^T$$

With $C_q$ the observations within the cluster $q$, $c_q$ the center of cluster $q$ and $c_E$ the center of $E$. The higher the index is, the denser and well separated the clusters are.

## 2. About these metrics

The three scores presented above are initially designed for convex clusters resulting from clustering techniques such as K-means. For Density-Based clustering such as HDBSCAN, the values of these scores will automatically seem "lower" (Moulavi et al. 2014). In the set-up that is about to be presented, this downward bias may be seen as an advantage as it might be interpreted as the minimum estimated quality of our clustering process.

### D.    A custom model to classify development finance activities

Now that we have investigated the motivation and origin of this work, we can delve into the new methodology deployed in itself, including the data cleaning and preprocessing and the model used.

### 1.    Data cleaning and preprocessing

One of the major challenges in this work is the treatment of 'raw data' (i.e. textual descriptions of projects) specifically the preprocessing of text data which involves translation, cleaning, tokenisation, and stemming. These steps are performed to reduce noise around the information encapsulated in the 'raw data' in order to enhance the quality and relevance of extracted features. Despite attempts to harmonize project description declarations, projects are declared heterogeneously depending on the declaring agency, their reporting habits, and the types of implemented projects. As the model used is from the family of transformers model (Vaswani et al. 2017), it is not necessary to realize any type of preprocessing to obtain a good understanding of the "raw data" variable. In fact, bi-directional transformer models are capable of contextualizing each document, allowing them to gain quality in natural language processing. Unlike previous models, they are less sensitive to noise and stop-words (e.g. "or", "and", "if", etc.), exploring the nuances embedded in these words. What sets transformers apart is their attention mechanism, which allows them to weigh the relevance of different parts of the input data when making predictions. This mechanism enables transformers to capture long-range dependencies in sequences, making them particularly effective for tasks involving sequential data like language translation, text generation, sentiment analysis, and more.

### 2.    Machine Learning Model used for Natural Language Processing

This section will be dedicated to the explanation of the global structure of the model implemented in this work. The global structure of this work can be illustrated by Figure 1.

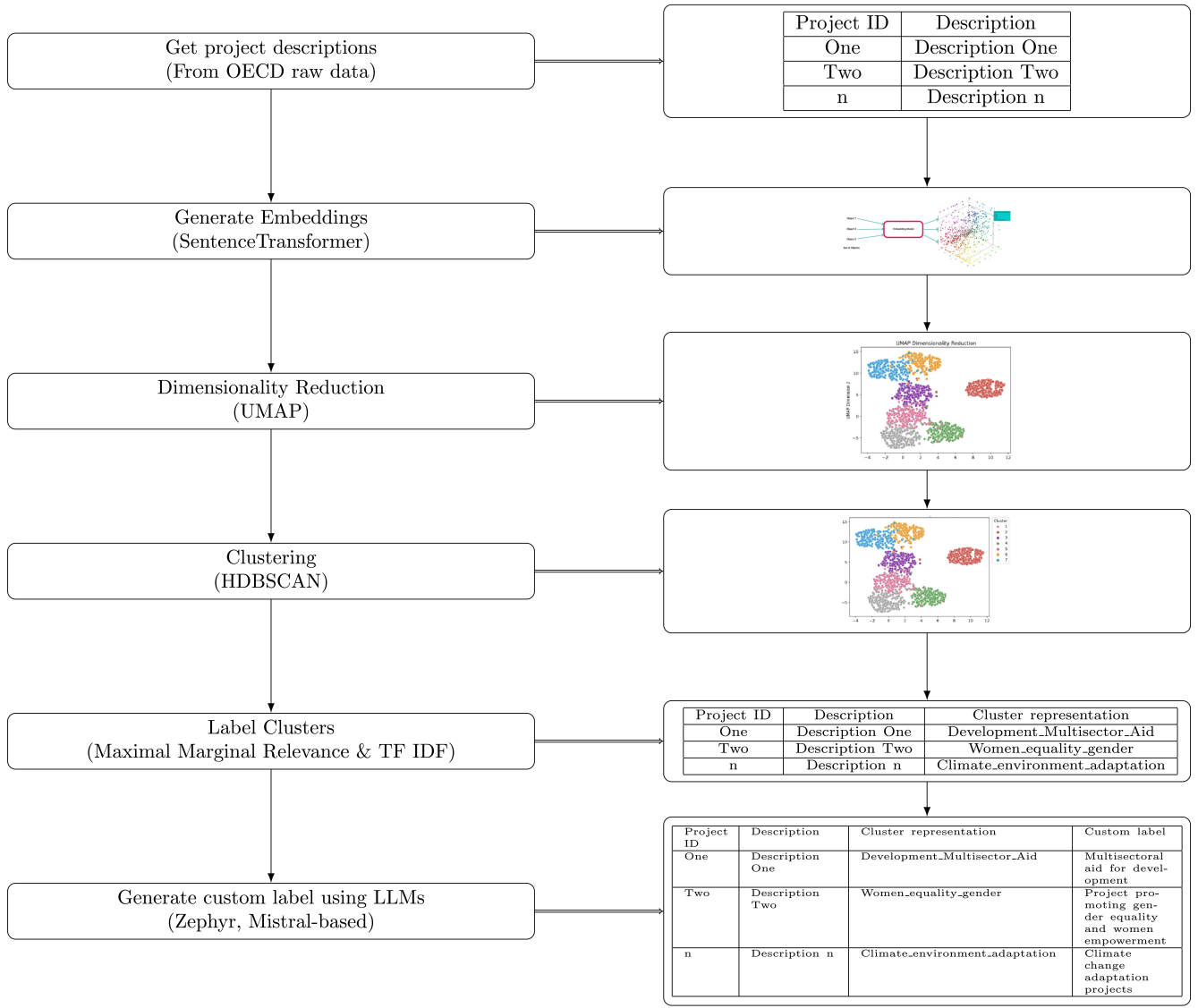

*Figure 1: Setup explanation*

As said earlier, an unsupervised learning approach is adopted. The main part of the process relies on the first step: text embedding, which is a multi-dimensional representation of words and their associations with other words. To do so, it is now common to use a BERT model (Devlin et al. 2018). Especially, the model used in this study is BERTopic (Grootendorst 2022), a BERT-based model specifically designed for quality clustering and topic modelling. BERTopic is divided in several steps detailed in Figure 2.

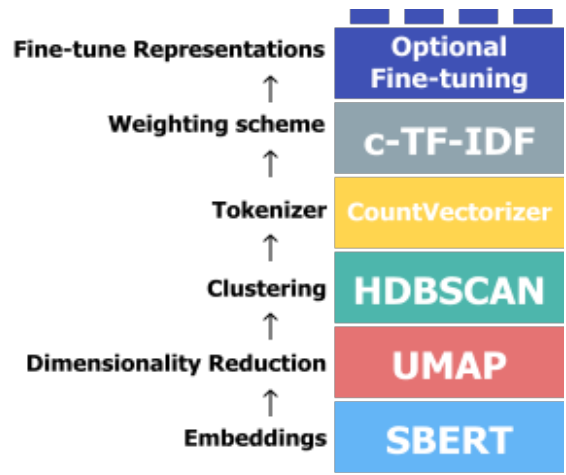

*Figure 2: BERTopic steps*

In explanations, The focus will be set on the four main parts of the process: **Embedding**, **Dimensionality reduction**, **Clustering** and **weighting schemes**.

*a)  Embedding*

An **embedding** is a relatively low-dimensional space into which you can translate high-dimensional vectors. Here, it is the vector representation of our raw data (project description) in a dimensional space. To do so, the model uses a transformers model (paraphrase-multilingual-MiniLM-L12-v2) from hugging face. It maps sentences and paragraphs to a 384-dimensional dense vector space. This specific model fits particularly to the set-up of data as it has been trained on 50 languages and project descriptions represent several languages including English, French, German or Dutch, as declarating agencies may declare projects in their official languages.

*b)  Dimensionality reduction*

In order to increase the computation speed of the model and to avoid the curse of dimensionality, a dimensionality reduction process is performed on original embeddings. In this setup, it has been chosen to use a quite new dimensionality reduction called Uniform Manifold Approximation and Projection (UMAP). UMAP is a dimension reduction technique that can be used for visualisation similarly to t-SNE, but also for general non-linear dimension reduction (McInnes, Healy, and Melville 2018). Dimensions of the embeddings are reduced to 12. UMAP has the advantage of preserving local and global "image" of the vectorised dataset. Other techniques could be performed such as principal component analysis, but then it will lose the overall structure of data and won't be interpretable in itself. The 12 dimensions choice has been made by a trade-off between keeping a higher level of information and gaining computational speed but also in efficiently estimating vector distances and similarity, as several distance metrics (e.g. Euclidean or cosine) do not perform well in high dimensional spaces (Xia et al. 2015).

*c)      Clustering process*

HDBSCAN (Hierarchical Density-Based Spatial Clustering of Applications with Noise, McInnes and Healy 2017) is used for clustering because of its strong performance on large datasets with high-dimensional embeddings. Another advantage of HDBSCAN over other clustering processes, such as K-Means, is that it supports non-convex vectorization of texts and finds the optimal number of clusters without any parameters, such as silhouette score, for example. The minimum cluster size is set at 500, meaning that each cluster contains at least 500 unique project descriptions to ensure a sufficient yet consistent number of projects in each cluster. One drawback concerning HDBSCAN is that it creates an outlier cluster in order to preserve the quality and "integrity" of each cluster. This drawback may not be an issue if the will is to identify and reveal a topic that surely exists in the dataset. Then, if one wants to work specifically on one or another topic and get the largest view of it, one possibility would be to use another process to classify all the remaining projects regarding those selected topics. The idea would be to take the projects present in the topics of interest as reference and then to compare it to projects from other clusters (including outliers projects). Then a machine learning algorithm could quite easily add to the clusters of interest projects that share a sufficient level of semantic similarity with those already present.

At the end, this new output will enable researchers to reveal hidden patterns in development project financing and implementation over the years, donors and recipient countries, refining the analysis of development finance.

*d)      Reducing outliers process*

As discussed in the Clustering Process part, the HDBSCAN is designed to get a high level of performance even with noise in targeted data. The main downside of this high performance is the creation of an "outlier cluster" that can be quite substantial in this case. However, one can re-run the algorithm to reduce the number of outliers. Indeed, each outlier project will be assigned to the nearest existing cluster based on some criteria. Those processes might be costly in terms of quality of the clustering. As we can see in Table 1, the impact of each outlier reduction step on the quality of the clustering is monitored.

*Table 1: Outliers reduction process*

| ClusteringScores | Original | After First Reduction | After Second Reduction |
|---|---|---|---|
| Silhouette Score | 0.72 | 0.58 | 0.46 |
| davies_bouldin Score | 0.42 | 1.55 | 1.84 |
| Calinski_harabasz Score | 234 463.20 | 20 956.50 | 21 697.72 |

It is worth noticing that HDBSCAN has a trade-off between high-quality clustering and a reduced number of outliers. Table 1 shows that each iteration of outliers reduction decrease the overall quality of the clustering based on the metrics described earlier.

*e)     Labeling clusters*

In order to get a representative label or title for the cluster automatically and instead of reading the project descriptions within a cluster and naming by hand, one can try to determine the importance or relevance score of each word from each description within a cluster, and comparatively to their relevance in other clusters. In the end, the description of clusters can be obtained with their most important and unique words. The model used is a Class-based TF-IDF procedure using scikit-learns TfidfTransformer as a base. For a term $x$ within class $c$:

$$W_{x,c} = \parallel \text{tf}_{x,c} \parallel \times \log\left(1 + \frac{A}{f_x}\right)$$

- $\text{tf}_{x,c}$ : frequency of word $x$ in class $c$

- $f_x$ : frequency of word $x$ across all classes

- $A$: average number of words per class

c-TF-IDF combines TF-IDF with the concept of multiple classes by treating each class as a single document. Initially, the term frequency (TF) for each word in each class is computed by summing up its occurrences and normalizing it. Then, this TF is multiplied by the Inverse Document Frequency (IDF), which is calculated as the logarithm of the average number of words per class divided by the frequency of the word across all classes.This allows to find the words that describe most the cluster regarding the others, rejecting the most common words in the entire dataset to account for each cluster unicity.

Once extracted the words that describes best each cluster and added set of the 5 most representative documents for each of it, one can use them in order to improve the labelling of clusters. To do so, a new large language model (LLM) is used. Derived from **Mistral**, Zephyr-7B-β is the second model in this series, and it is a fine-tuned version of Mistral-7B-v0.1. A prompt for this LLM is redacted to generate appropriate names for found clusters based on their top ten words and five representative documents (project descriptions that accurately characterize each cluster).

## III.   Results

As depicted earlier, the main output of this work is a new and rearranged dataset with a thinner level of development project classification, determined by unsupervised machine learning techniques. The proposed methodology, described in Figure 1, provides several practical implications for policymakers, economists, and stakeholders involved in development planning. Automated categorisation enables rapid analysis, leading to more informed decision-making and resource allocation. It also enables a more detailed analysis of development projects with finer categories, leveraging the study of several finance for development topics such as climate finance or poverty reduction. This new classification may reveal hidden patterns in project classifications between and within donors and implementing agencies. It can be compared with already existing sectors and thematic variables for a more diverse analysis. Another advantage of this work is that it breaks down the available data at a project level, allowing for reclassification and overcoming the micro/macroeconomics dichotomy in empirical analysis.

### A.   High level of detail clustering

The primary outcome of this work is a novel classification encompassing 406 distinct topics of the projects listed in the OECD CRS dataset. To conserve the integrity of clusters, the model used for clustering creates outlier projects. For transparency purposes, Figure 3 shows what represent the outliers in the final dataset. It can be observed that this clustering process is able to capture more than half of the development projects since their monitoring by the OECD. Furthermore, this algorithm is able to regroup projects with high semantic similarity and then extract and label the common theme of these newly created groups.

| Topic | CRS dataset after restraining to one similar project description | Attrition of clustering process (%) | Complete CRS dataset | Attrition of clustering process (%) |
|---|---|---|---|---|
| Total of outliers obs. | 861,793 | 47.2 | 2,457,737 | 48.4 |
| Total obs. clustered | 964,170 | 52.8 | 2,615,421 | 51.6 |
| Total | 1,825,963 | 100.0 | 5,073,158 | 100.0 |

## Distribution of tracked and untracked project per year

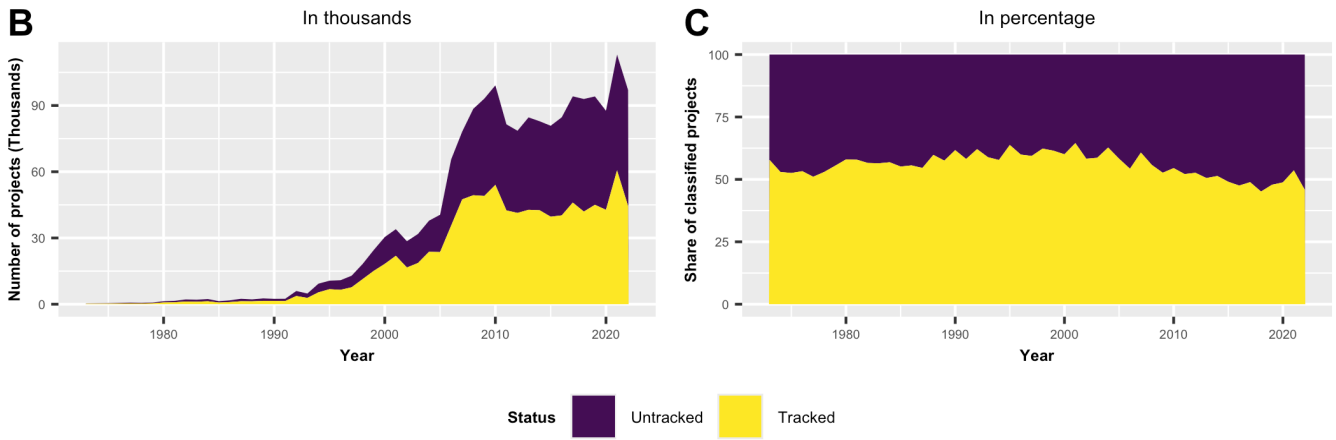

*Figure 3: Tracked and untracked (outliers) projects*

Figure 3 also shows that tracked and untracked projects follow the same trend across time and that the proportion of projects classified as outliers remain quite stable throughout the years. It suggests that the clustering process does not suffer from any time-varying effect.

It shall be mentioned that the algorithm has been re-run two times to reduce the number of outliers. To do so, the algorithm took outlier projects and assigned them in the existing clusters based on similarities with already assigned projects. Thus, this process may not be independent of the final quality of clustering (as seen in Table 1.

### B.    Comparison with the OECD nomenclature

We can compare this clustering and topic modelling process to the original project classification from the OECD CRS dataset available in Table 3 in the extended data to see the revealed patterns from our topic modeling methodology. We can already observe that whereas the OECD purpose codes stand for 234 categories, our clustering process is able to identify 406 distinct topics. Table 2 shows us the difference of level of detail per sector. Each "●" accounts for a number of purpose codes or topics that are in both classifications whereas each "○" accounts for an additional cluster compared to the number of purpose codes in each sector.

| OECD Sector Name | Representation |
| --- | --- |
| Administrative Costs of Donors | ●○○○○ |
| I.1.a. Education, Level Unspecified | ●●○○○○ |
| I.1.b. Basic Education | ●●○○○○○○○○ |
| I.1.c. Secondary Education | ●● |
| I.1.d. Post-Secondary Education | ●●○○○○○○○○○○○○ |
| I.2.a. Health, General | ●●●●○○○ |
| I.2.b. Basic Health | ●●●●●●●○○○○○○○○ ○○ |
| I.3. Population Policies/Programmes & Reproductive Health | ●●●●○○○○○○○○○○○ |
| I.4. Water Supply & Sanitation | ●●●●○○○○○○ |
| I.5.a. Government & Civil Society-general | ●●●●●●●●●●●○○○ ○○○○○○○○○○○○○○ ○○○○○○○○○○○○○○ ○○○○○○○○○○○○○○ ○○○○○○○○○○○○○○ ○○○○○○○○○○○○ |
| I.5.b. Conflict, Peace & Security | ●●●●●○○○○○○ |
| I.6. Other Social Infrastructure & Services | ●●●●●●●●●●○○○○○ ○○○○○○○○○○○○○○ ○○○○○ |
| II.1. Transport & Storage | ●●●●●○○○○○○ |
| II.2. Communications | ●●●● |
| II.3.a. Energy Policy | ●○ |
| II.3.b. Energy generation, renewable sources | ●●●●●● |
| II.3.c. Energy generation, non-renewable sources | ● |
| II.3.e. Nuclear energy plants | ● |
| II.3.f. Energy distribution | ●○ |
| II.4. Banking & Financial Services | ●●○○○○○ |
| II.5. Business & Other Services | ●○○ |
| III.1.a. Agriculture | ●●●●●●●●●●●○○○○ ○○○○○ |
| III.1.b. Forestry | ● |
| III.1.c. Fishing | ● |
| III.2.a. Industry | ●●●● |
| III.2.b. Mineral Resources & Mining | ●●○ |
| III.3.a. Trade Policies & Regulations | ●● |
| III.3.b. Tourism | ● |
| IV.1. General Environment Protection | ●●●●●●○○○○○○○○ ○○○○○○ |
| IV.2. Other Multisector | ●●●●●●○○○○○○○○ ○○○○○○ |
| IX. Unallocated / Unspecified | ●●○○○○○○○○○○○○ ○ |
| Refugees in Donor Countries | ●○ |
| VI.1. General Budget Support | ● |

| OECD Sector Name | Representation |
|---|---|
| VI.2. Development Food Assistance | ●○○○○○○○○ |
| VI.3. Other Commodity Assistance | ● |
| VII. Action Relating to Debt | ●○ |
| VIII.1. Emergency Response | ●●○○○○○○○○○○○○○ ○○○○○○○○○○ |
| VIII.2. Reconstruction Relief & Rehabilitation | ●○○ |
| VIII.3. Disaster Prevention & Preparedness | ●○ |

● represents the number of Topics and PurposeCodes in common. ○ represents additional Topics beyond the number of PurposeCodes. Each Topic is assigned to a Sector based on the number of projects that are monitored in purpose codes for each topic. For example, the topic Health will be assigned to sector I.2.a. Health, General if the majority of the projects clustered in Health are monitored with Purpose Codes present in the Sector I.2.a. Health, General. This table summarizes the distribution of Topics and PurposeCodes across different Sectors.

Table 2 illustrates that each category already present in the CRS methodology is represented at least once in this new methodology, highlighting the algorithm's ability to retain all initial information while adding more details in the classification. It also shows some disparities in terms of breaking down some categories. This phenomenon may suggest that highly specific categories are more likely to be fully represented within themselves, whereas more general or "vague" categories (e.g., I.5.a. Government & Civil Society-general) tend to be divided into multiple clusters. This division highlights the granularity of the algorithm.

In fact, this application of machine learning illustrates the ability of artificial intelligence to provide a thinner and clearer understanding of our societies. Thanks to this work, researchers and policymakers are able to analyse development aid for LGBTQ rights, or development finance for beekeeping and honey production, as distinct issues within overall development finance, given the existence of clusters related to these same issues. This application of machine learning techniques allows for a better understanding of trends and fashions in global development finance, examining the emergence and decline of certain themes over time. The following section will discuss further of the potential impact of this work in terms of its applications and limitations.

## IV.    Discussion

This application of machine learning, especially natural language processing, may induce several possible applications in monitoring global development finance, including the following examples. Yet, the proposed method includes limitations one should consider.

## A. Possible applications

As explained earlier, the main goal of this work is to provide valuable insights concerning development finance and aid patterns and to leverage the possibilities of studies concerning development finance topic, going from very specific to larger ones. The HDBSCAN approach, which will be described in the methods section, guarantees this modularity thanks to its hierarchical clustering technique, allowing aggregation of clusters at different levels for more global subjects such as global climate or health financing. It also allows micro-level analysis of aid such as studying the great apes' conservation projects presented in Figure 6. The HDBSCAN model, as it is suited for high-density data setup, also allows and even fosters to use project-level data, enabling in the case of this work dyadic analysis as studying the Syrian emergency Aid flows as displayed in Figure 7. Last, the topic modelling process allows to study trends in Aid within differents topics. To illustrate, these differents scenario will be developed in this section.

### 1. A donor analysis application: climate finance-related activities

To illustrate this purpose, we can move on to testing its implications for one or several development themes. Let's assume we are researchers in climate economics and we want to have an idea of global climate finance. As a policymaker or researcher assessing the overall level of development finance committed and disbursed for climate and environmental purposes, it may be necessary to select only projects dealing with those themes. Figure 4 in extended data shows an example of the number of unique project descriptions from the climate and environment-related cluster. The importance of these descriptions can be interactively determined over time on the following website: https://pierrebeaucoral.github.io/project/crs-ml/topics_over_time_visualization.html.

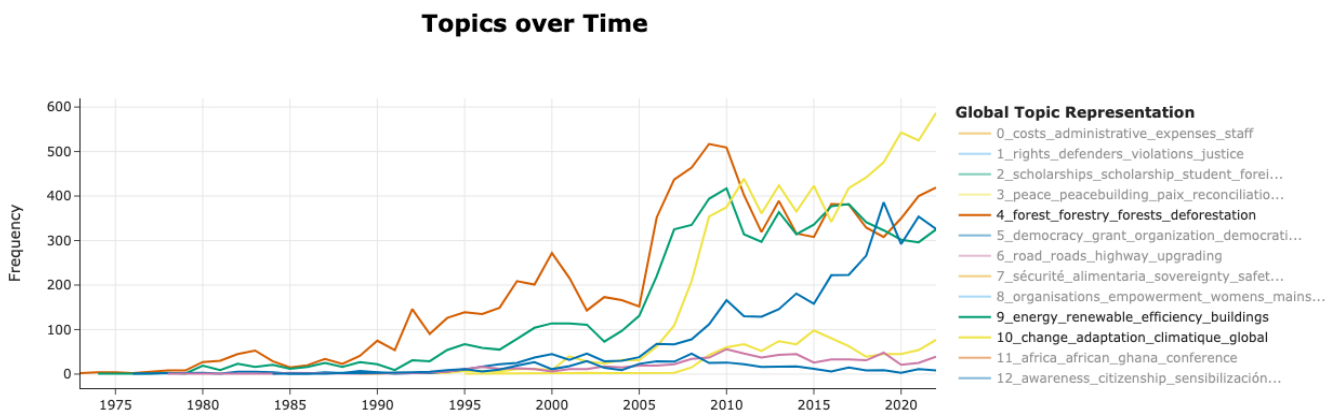

*Figure 4: Example for climate-related cluster from website*

If we agglomerate the amounts committed and disbursed from projects sharing the descriptions coming from those selected clusters, we get an estimation of global climate finance in Figure 9 (in extended data). Now, if we compare these amounts to those coming from the OECD CRS Rio markers in Figure 5, we can observe that the level of disbursed and commited flows are different based on textual descriptions compared to self declared binary variables. Figure 5 shows us a difference between 30 and 40 billions constant US dollars from textual analysis and binary variables such as Rio markers.

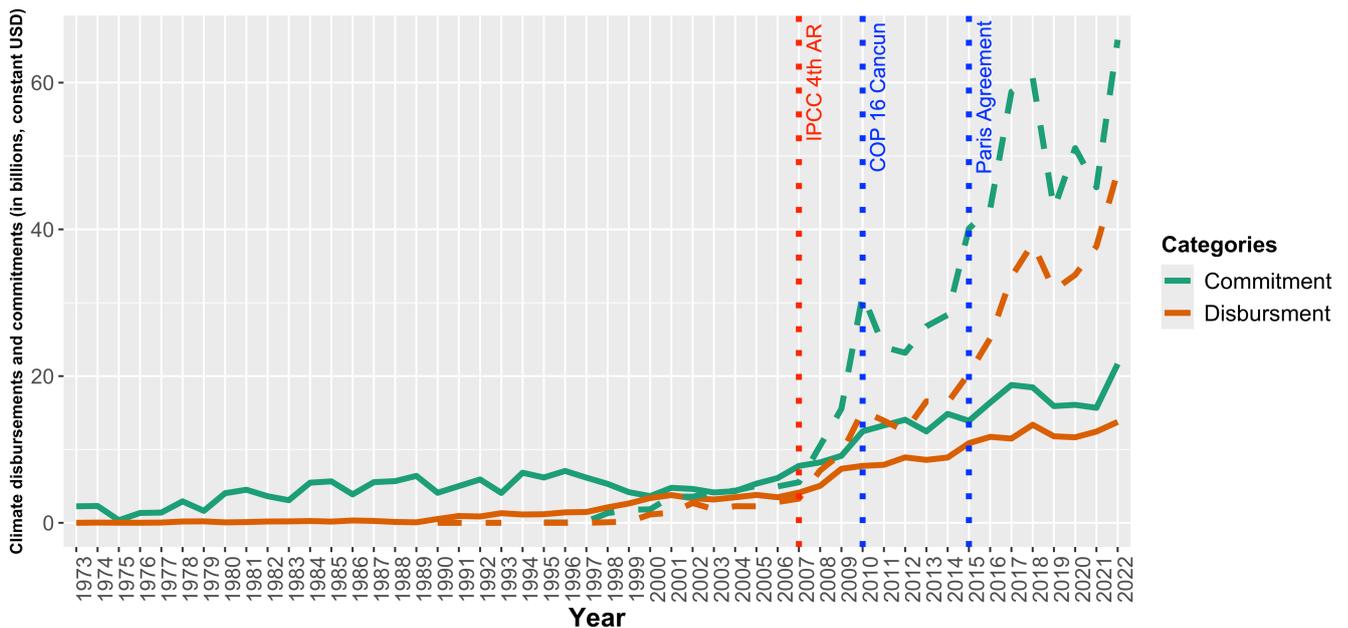

*The dashed line represents the CRS estimation of climate finance according to Rio markers*

*Figure 5: Evolution of climate finance (CRS vs Machine Learning)*

Figure 5 displays potential differences in project selection between those with self-declared climate or environmental objectives that are only illustrated by a dummy variable, which may suffer from over-reporting and moral hazard between implementing agencies and other development actors, and a technique relying on project descriptions, a more qualitative variable that may contain more information regarding development activities. Here, we seem to get a huge difference regarding level of commited and disbursed amount but also on the global trend of those financing. In this case, the use of artificial intelligence enables to get a more critical look into donor declaration regarding their financing goals, such as adressing climate change.

## 2.    A recipient analysis application: Great Apes conservation development projects

In this second application, one could desire to study the effects of dollars spent in one or another "niche" or very specific field of development finance. Then it will be needed to identify a specific topic and the areas where projects from this topic are located. Here, the application selected is the "Great Apes Conservation Parks" Cluster. Then one can map the location of these projects according to the number of projects per countries in order to get an idea of the area of study. The results of such idea can be highlighted in Figure 6.

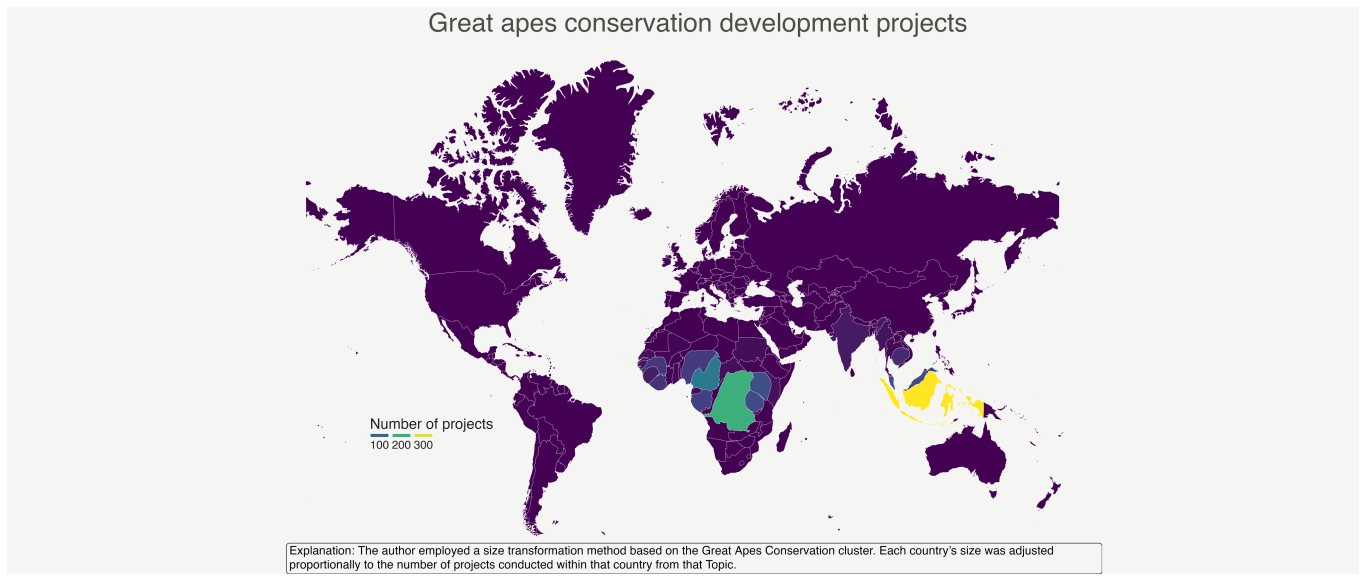

*Figure 6: Extraction of projects coming from the great apes conservation cluster*

According to figure 6, next step should be to make a choice regarding studying Central, West Africa or Indonesia for example.

### 3.    A dyadic analysis application: Studying Syrian humanitarian aid flows

Then, one can also want to study relationship between donor and recipient countries for specific topics. For example, we may wonder what could be the relative importance of a topic for donors. For example, would the donor countries closest to Syria be encouraged to provide more support?

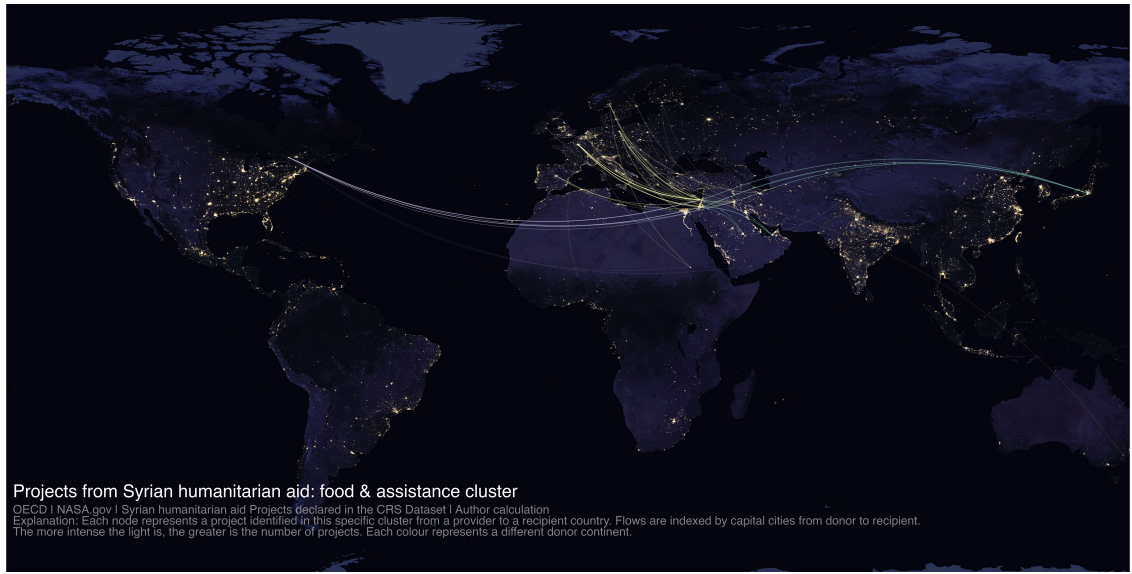

*Figure 7: Extraction of projects coming from the Syrian humanitarian aid cluster*

The figure 7 shows the efforts of donors regarding Syrian humanitarian aid, the destination of this aid, and the "strength" of the relationship between donor and recipient countries. The strength is illustrated by the brightness of the link, the brighter the link, the greater the number of projects. It is likely that distance from Syria influences the sending and receiving of aid. For example, Australian aid seems insignificant compared to European aid. However, it may not be the only determinant of Syrian crisis aid, as we can see that the United States and Japan are also both represented by bright links.

## 4.    Is aid a fashion victim?

Aid represents a complex and multifaceted subject that cannot be adequately captured through simplistic sectoral categorizations. Such oversimplification risks obscuring the nuanced objectives of Aid and may lead to a generalized portrayal that fails to recognize instances where Aid extends beyond its traditional mandates—such as eradicating hunger, fostering peace, and alleviating poverty—toward more contemporary and trend-driven goals.

# Is aid a fashion victim?

Analysis of diverse topic repartition in declared development projects
CRS OECD dataset

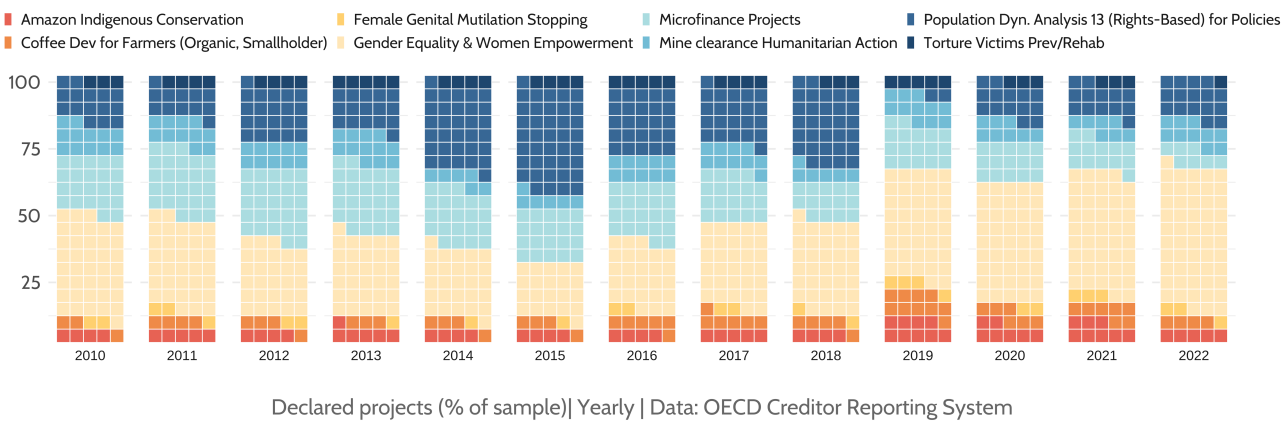

Declared projects (% of sample)| Yearly | Data: OECD Creditor Reporting System

*Figure 8: Topic trend analysis: An example.*

Figure 8 provides an initial insight, highlighting that certain topics appear to hold greater significance for donors or recipients than others. For instance, in the last ten years, the development of coffee production may generate a comparable number of projects annually to those aimed at conserving Indigenous peoples in the Amazon. Figure 8 also provide indications that traditional development topics, such as microfinance, may be decreasing in prominence over time compared to other emerging areas of focus. These insights may prompt reflection on the alignment of development agencies' agendas with their original mandates.

## B.    Policy implications

The application of advanced machine learning techniques like BERTopic enables policymakers and researchers to gain a granular understanding of development finance. This can help identify discrepancies or inefficiencies in aid allocation and promote transparency by categorizing projects based on detailed narrative descriptions rather than self-declared objectives. Such transparency is critical for maintaining public trust in development finance initiatives. By revealing hidden thematic clusters, this methodology aids in pinpointing underfunded or emerging areas within development finance. For instance, identifying niche topics such as LGBTQ rights or conservation projects can lead to more targeted and effective resource allocation, ensuring aid reaches sectors and populations with specific needs. The findings highlight significant differences between donor-reported metrics (e.g., Rio markers) and the classification derived from narrative descriptions. This can serve as a tool to verify donor commitments and combat over-reporting, thus fostering accountability and adherence to international agreements, such as the Paris Agreement. Governments and development agencies can use these insights to align their strategies with evidence-based priorities. The dynamic clustering approach facilitates real-time monitoring of trends, enabling agile responses to emerging global challenges like pandemics or climate crises. By providing detailed, project-level data, this approach supports localized planning and implementation. For instance, identifying clusters related to specific geographic regions or themes can aid in customizing solutions to local challenges, thus improving development outcomes.

The study demonstrates the potential to refine climate finance tracking by focusing on project narratives. This approach could uncover additional funding streams or unreported contributions to climate mitigation and adaptation, offering a more comprehensive view of global climate finance flows. The refined classification of development projects enables better comparisons across donors and recipients, aiding in international benchmarking. Policymakers can use these insights to evaluate the effectiveness of different aid strategies and identify best practices. Multilateral organizations like the OECD could integrate this methodology to complement existing classification systems, such as purpose codes and Rio markers. This would enhance the global development finance architecture and support efforts toward achieving the Sustainable Development Goals (SDGs).

### C.    Limitations

Even though this work deploys state of the art methodologies to get a better description of the topics of interest in development finance strategies across time, it might show some limitations concerning the "raw material" of this analysis, the methodology use and the insights of this work in terms of volume of allocated aid.

The quality of textual description of projects has improved across the period of this study, however it remains heterogeneous in terms of length and details between implementing agencies and across time. Even by using a performing multilingual BERT model for embedding, this heterogeneity may reduce the quality of embedding and then the ability of precise clustering by the HDBSCAN model. One way of addressing this difficulty would be to compute the same analysis by getting the project descriptions available on the donors websites. It will improve the quality of embedding, yet it will be more time consuming and less exhaustive, relating only on publicly available project descriptions today.

Concerning the methodology of this study, one main drawback shall be the production of outliers during the clustering process. Still, this outlier production is done for several purposes which are (i) preserving the coherence of a topic/cluster; (ii) being able to identify very specific subject of analysis that would be drowned in the mass of other more "generic" description. One possible way of addressing this issue would be to re-run a supervised machine learning to assign to your build sample outlier projects that are close to your sample project descriptions in terms of textual similarities.

About insights in terms of allocation of aid in volume, because the clustering process might be considered as conservative, it should be seen in itself as a "lower bound" of global amount committed or disbursed in a specific cluster or topic. If one take back the case study on climate finance provided earlier, the truth might be in between the algorithm estimation and the one using OECD Rio Markers. One way of getting a more precise estimation of a specific topic would be to explore the suggested methodology in the above paragraph by re-running a supervised machine learning technique. In case of climate finance, one can use climatefinanceBERT (Toetzke, Stünzi, and Egli 2022), with the code available at the following link: https://github.com/MalteToetzke/consistent-and-replicable-estimation-of-bilateral-climate-finance.

### D.    Concluding remarks

This work presents an innovative application of machine learning and natural language processing techniques to categorise development projects based on narrative descriptions. The use of the OECD CRS dataset, which is a well-known reference in development finance data sources, demonstrates the feasibility and effectiveness of the proposed approach. The findings contribute to the ongoing discourse on innovative techniques in economics, machine learning, and project analysis. This work also allows for a better understanding of all development assistance, from its geographical to its thematic distribution. Although this work provides valuable insights, its methodology can also be applied in other fields of science. The use of text embeddings and classification methods can be of interest to several fields, including behavioural sciences for sentiment analysis, labour and international trades and geopolitics studies for contract classification, as well as historical or human sciences research for creating quantitative datasets from texts and other qualitative inputs. Overall, these methods, including text embeddings, classification, and topic modelling, are crucial for uncovering hidden patterns in academic research, particularly through meta-analysis. The following section will detail the methodology employed in this study.

# VI. Data availability statement

Raw data are available on the OECD website: https://data-explorer.oecd.org/vis?tm=Creditor%20Reporting%20System&pg=0&snb=7&df[ds]=dsDisseminateFinalDMZ&df[id]=DSD_CRS%40DF_CRS&df[ag]=OECD.DCD.FSD&df[vs]=1.1&dq=DAC..1000.100._T_.T.D.Q._T..&lom=LASTNPERIODS&lo=5&to[TIME_PERIOD]=false, transformed data and results of the analysis is obtainable through replication with the scripts available here: https://github.com/PierreBeaucoral/ML-clustering-of-development-activities.

# VII. Extended Data

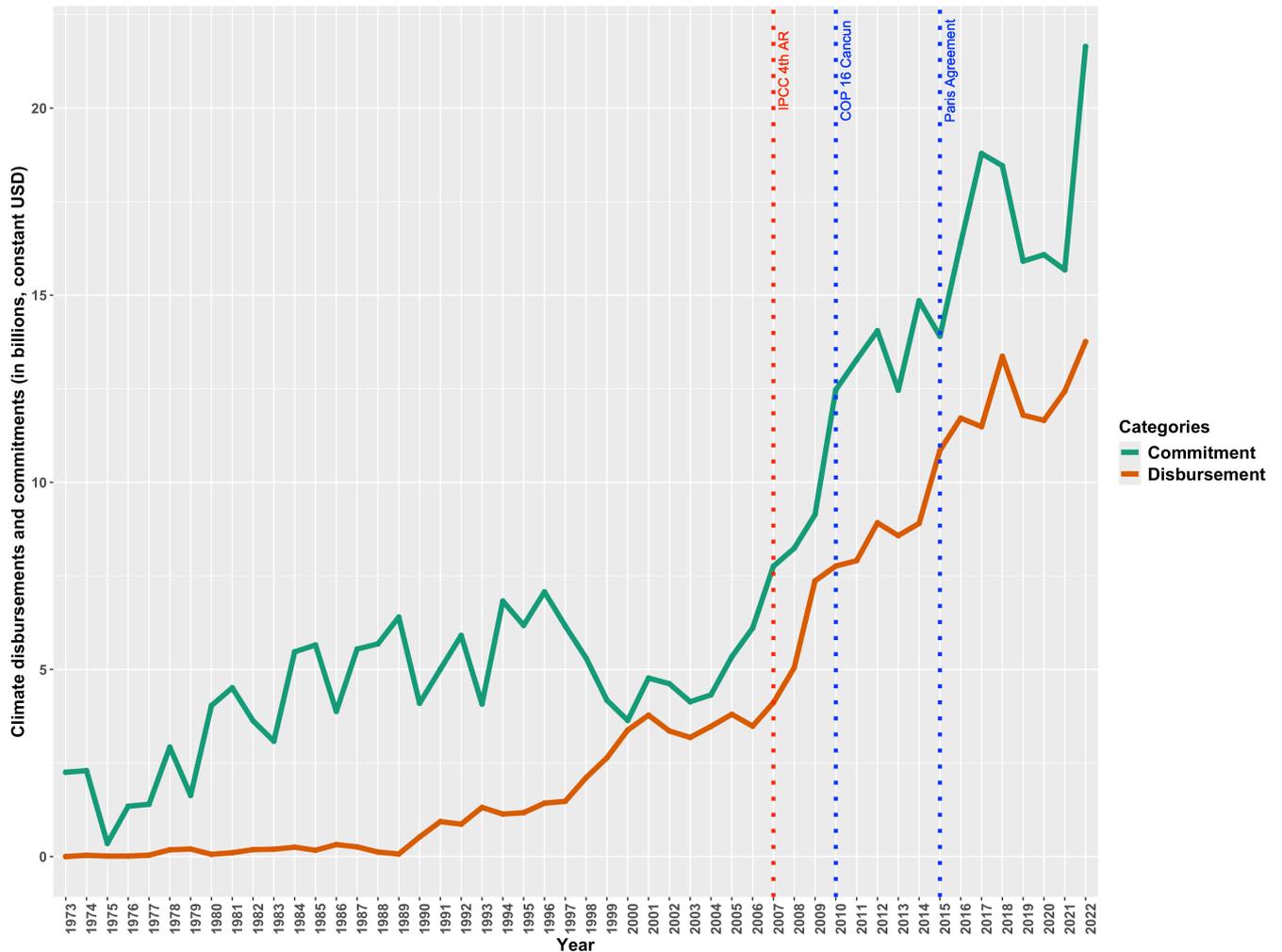

*Estimation of global finance levels declared in the CRS dataset thanks to the clustering process*

*Figure 9: Example of estimation of global climate finance*

# A. OECD purpose codes

*Table 3: OECD purpose code classification*

| DAC 5 | CRS | DESCRIPTION | Clarifications / Additional notes on coverage |
|-------|-----|-------------|------------------------------------------------|
| 110 | | Education | |
| 111 | | Education, Level Unspecified | The codes in this category are to be used only when level of education is unspecified or unknown (e.g. training of primary school teachers should be coded under 11220). |
| | 11110 | Education policy and administrative management | Education sector policy, planning and programmes; aid to education ministries, administration and management systems; institution capacity building and advice; school management and governance; curriculum and materials development; unspecified education activities. |
| | 11120 | Education facilities and training | Educational buildings, equipment, materials; subsidiary services to education (boarding facilities, staff housing); language training; colloquia, seminars, lectures, etc. |
| | 11130 | Teacher training | Teacher education (where the level of education is unspecified); in-service and pre-service training; materials development. |
| | 11182 | Educational research | Research and studies on education effectiveness, relevance and quality; systematic evaluation and monitoring. |
| 112 | | Basic Education | |
| | 11220 | Primary education | Formal and non-formal primary education for children; all elementary and first cycle systematic instruction; provision of learning materials. |
| | 11230 | Basic life skills for adults | Formal and non-formal education for basic life skills for adults (adults education); literacy and numeracy training. Excludes health education (12261) and activities related to prevention of noncommunicable diseases. (123xx). |
| | 11231 | Basic life skills for youth | Formal and non-formal education for basic life skills for young people. |
| | 11232 | Primary education equivalent for adults | Formal primary education for adults. |
| | 11240 | Early childhood education | Formal and non-formal pre-school education. |
| | 11250 | School feeding | Provision of meals or snacks at school; other uses of food for the achievement of educational outcomes including 'take-home' food rations provided as economic incentives to families (or foster families, or other child care institutions) in return for a child's regular attendance at school; food provided to adults or youth who attend literacy or vocational training programmes; food for pre-school activities with an educational component. These activities may help reduce children's hunger during the school day if provision of food/meals contains bioavailable nutrients to address specific nutrition needs and have nutrition expected outcomes in school children, or if the rationale mainstream nutrition or expected outcome is nutrition-linked. |
| | 11260 | Lower secondary education | Second cycle systematic instruction at junior level. |
| 113 | | Secondary Education | |
| | 11320 | Upper Secondary Education (modified and includes data from 11322) | Second cycle systematic instruction at senior levels. |
| | 11330 | Vocational training | Elementary vocational training and secondary level technical education; on-the job training; apprenticeships; including informal vocational training. |
| 114 | | Post-Secondary Education | |

| DAC 5 | CRS | DESCRIPTION | Clarifications / Additional notes on coverage |
|---|---|---|---|
| | 11420 | Higher education | Degree and diploma programmes at universities, colleges and polytechnics; scholarships. |
| | 11430 | Advanced technical and managerial training | Professional-level vocational training programmes and in-service training. |
| 120 | | Health | |
| 121 | | Health, General | |
| | 12110 | Health policy and administrative management | Health sector policy, planning and programmes; aid to health ministries, public health administration; institution capacity building and advice; medical insurance programmes; including health system strengthening and health governance; unspecified health activities. |
| | | Health statistics and data | Collection, production, management and dissemination of statistics and data related to health. Includes health surveys, establishment of health databases, data collection on epidemics, etc. |
| | 12181 | Medical education/training | Medical education and training for tertiary level services. |
| | 12182 | Medical research | General medical research (excluding basic health research and research for prevention and control of NCDs (12382)). |
| | 12191 | Medical services | Laboratories, specialised clinics and hospitals (including equipment and supplies); ambulances; dental services; medical rehabilitation. Excludes noncommunicable diseases (123xx). |
| 122 | | Basic Health | |
| | 12220 | Basic health care | Basic and primary health care programmes; paramedical and nursing care programmes; supply of drugs, medicines and vaccines related to basic health care; activities aimed at achieving universal health coverage. |
| | 12230 | Basic health infrastructure | District-level hospitals, clinics and dispensaries and related medical equipment; excluding specialised hospitals and clinics (12191). |
| | 12240 | Basic nutrition | Micronutrient deficiency identification and supplementation; Infant and young child feeding promotion including exclusive breastfeeding; Non-emergency management of acute malnutrition and other targeted feeding programs (including complementary feeding); Staple food fortification including salt iodization; Nutritional status monitoring and national nutrition surveillance; Research, capacity building, policy development, monitoring and evaluation in support of these interventions. Use code 11250 for school feeding and 43072 for household food security. |
| | 12250 | Infectious disease control | Immunisation; prevention and control of infectious and parasite diseases, except malaria (12262), tuberculosis (12263), COVID-19 (12264), HIV/AIDS and other STDs (13040). It includes diarrheal diseases, vector-borne diseases (e.g. river blindness and guinea worm), viral diseases, mycosis, helminthiasis, zoonosis, diseases by other bacteria and viruses, pediculosis, etc. |
| | 12261 | Health education | Information, education and training of the population for improving health knowledge and practices; public health and awareness campaigns; promotion of improved personal hygiene practices, including use of sanitation facilities and handwashing with soap. |
| | 12262 | Malaria control | Prevention and control of malaria. |
| | 12263 | Tuberculosis control | Immunisation, prevention and control of tuberculosis. |
| | 12264 | COVID-19 control | All activities related to COVID-19 control e.g. information, education and communication; testing; prevention; immunisation, treatment, care. |
| | 12281 | Health personnel development | Training of health staff for basic health care services. |

| DAC 5 | CRS | DESCRIPTION | Clarifications / Additional notes on coverage |
|-------|-----|-------------|-----------------------------------------------|
| 123 | | Non-communicable diseases (NCDs) | |
| | 12310 | NCDs control, general | Programmes for the prevention and control of NCDs which cannot be broken down into the codes below. |
| | 12320 | Tobacco use control | Population/individual measures and interventions to reduce all forms of tobacco use in any form. Includes activities related to the implementation of the WHO Framework Convention on Tobacco Control, including specific high-impact demand reduction measures for effective tobacco control. |
| | 12330 | Control of harmful use of alcohol and drugs | Prevention and reduction of harmful use of alcohol and psychoactive drugs; development, implementation, monitoring and evaluation of prevention and treatment strategies, programmes and interventions; early identification and management of health conditions caused by use of alcohol and drugs [excluding narcotics traffic control (16063)]. |
| | 12340 | Promotion of mental health and well-being | Promotion of programmes and interventions which support mental health and well-being resiliency; prevention, care and support to individuals vulnerable to suicide. Excluding treatment of addiction to tobacco, alcohol and drugs (included in codes 12320 and 12330). |
| | 12350 | Other prevention and treatment of NCDs | Population/individual measures to reduce exposure to unhealthy diets and physical inactivity and to strengthen capacity for prevention, early detection, treatment and sustained management of NCDs including: Cardiovascular disease control: Prevention, screening and treatment of cardiovascular diseases (including hypertension, hyperlipidaemia, ischaemic heart diseases, stroke, rheumatic heart disease, congenital heart disease, heart failure, etc.). Diabetes control: Prevention, screening, diagnosis, treatment and management of complications from all types of diabetes. Exposure to physical inactivity: Promotion of physical activity through supportive built environment (urban design, transport), sports, health care, schools and community programmes and mass media campaign. Exposure to unhealthy diet: Programmes and interventions that promote healthy diet through reduced consumption of salt, sugar and fats and increased consumption of fruits and vegetables e.g. food reformulation, nutrient labelling, food taxes, marketing restriction on unhealthy foods, nutrition education and counselling, and settings-based interventions (schools, workplaces, villages, communities). Cancer control: Prevention (including immunisation, HPV and HBV), early diagnosis (including pathology), screening, treatment (e.g. radiotherapy, chemotherapy, surgery) and palliative care for all types of cancers. Implementation, maintenance and improvement of cancer registries are also included. Chronic respiratory diseases: Prevention, early diagnosis and treatment of chronic respiratory diseases, including asthma. Excludes: Tobacco use control (12320), Control of harmful use of alcohol and drugs (12330), research for the prevention and control of NCDs (12382). |
| | 12382 | Research for prevention and control of NCDs | Research to enhance understanding of NCDs, their risk factors, epidemiology, social determinants and economic impact; translational and implementation research to enhance operationalisation of cost-effective strategies to prevent and control NCDs; surveillance and monitoring of NCD mortality, morbidity, risk factor exposures, and national capacity to prevent and control NCDs. |
| 130 | | Population Policies/Programmes & Reproductive Health | |
| | 13010 | Population policy and administrative management | Population/development policies; demographic research/analysis; reproductive health research; unspecified population activities. (Use purpose code 15190 for data on migration and refugees. Use code |

| DAC 5 | CRS | DESCRIPTION | Clarifications / Additional notes on coverage |
|-------|-----|-------------|-----------------------------------------------|
| | | | 13096 for census work, vital registration and migration data collection.) |
| | | Population statistics and data | Collection, production, management and dissemination of statistics and data related to Population and Reproductive Health. Includes census work, vital registration, migration data collection, demographic data, etc. |
| | 13020 | Reproductive health care | Promotion of reproductive health; prenatal and postnatal care including delivery; prevention and treatment of infertility; prevention and management of consequences of abortion; safe motherhood activities. |
| | 13030 | Family planning | Family planning services including counselling; information, education and communication (IEC) activities; delivery of contraceptives; capacity building and training. |
| | 13040 | STD control including HIV/AIDS | All activities related to sexually transmitted diseases and HIV/AIDS control e.g. information, education and communication; testing; prevention; treatment, care. |
| | 13081 | Personnel development for population and reproductive health | Education and training of health staff for population and reproductive health care services. |
| 140 | | Water Supply & Sanitation | |
| | 14010 | Water sector policy and administrative management | Water sector policy and governance, including legislation, regulation, planning and management as well as transboundary management of water; institutional capacity development; activities supporting the Integrated Water Resource Management approach (IWRM: see box below). |
| | 14015 | Water resources conservation (including data collection) | Collection and usage of quantitative and qualitative data on water resources; creation and sharing of water knowledge; conservation and rehabilitation of inland surface waters (rivers, lakes etc.), ground water and coastal waters; prevention of water contamination. |
| | 14020 | Water supply and sanitation - large systems | Programmes where components according to 14021 and 14022 cannot be identified. When components are known, they should individually be reported under their respective purpose codes: water supply [14021], sanitation [14022], and hygiene [12261]. |
| | 14021 | Water supply - large systems | Potable water treatment plants; intake works; storage; water supply pumping stations; large scale transmission / conveyance and distribution systems. |
| | 14022 | Sanitation - large systems | Large scale sewerage including trunk sewers and sewage pumping stations; domestic and industrial waste water treatment plants. |
| | 14030 | Basic drinking water supply and basic sanitation | Programmes where components according to 14031 and 14032 cannot be identified. When components are known, they should individually be reported under their respective purpose codes: water supply [14031], sanitation [14032], and hygiene [12261]. |
| | 14031 | Basic drinking water supply | Rural water supply schemes using handpumps, spring catchments, gravity-fed systems, rainwater collection and fog harvesting, storage tanks, small distribution systems typically with shared connections/points of use. Urban schemes using handpumps and local neighbourhood networks including those with shared connections. |
| | 14032 | Basic sanitation | Latrines, on-site disposal and alternative sanitation systems, including the promotion of household and community investments in the construction of these facilities. (Use code 12261 for activities promoting improved personal hygiene practices.) |
| | 14040 | River basins development | Infrastructure-focused integrated river basin projects and related institutional activities; river flow control; dams and reservoirs |

| DAC 5 | CRS | DESCRIPTION | Clarifications / Additional notes on coverage |
|---|---|---|---|
| | | | [excluding dams primarily for irrigation (31140) and hydropower (23220) and activities related to river transport (21040)]. |
| | 14050 | Waste management/disposal | Municipal and industrial solid waste management, including hazardous and toxic waste; collection, disposal and treatment; landfill areas; composting and reuse. |
| | 14081 | Education and training in water supply and sanitation | Education and training for sector professionals and service providers. |
| 150 | | Government & Civil Society | |
| 151 | | Government & Civil Society-general | N.B. Use code 51010 for general budget support. |
| | 15110 | Public sector policy and administrative management | Institution-building assistance to strengthen core public sector management systems and capacities. This includes general public policy management, co-ordination, planning and reform; human resource management; organisational development; civil service reform; e-government; development planning, monitoring and evaluation; support to ministries involved in aid co-ordination; other ministries and government departments when sector cannot be specified. (Use specific sector codes for development of systems and capacities in sector ministries. For macro-economic policy use code 15142. For public procurement use code 15125.) |
| | | Foreign affairs | Administration of external affairs and services. |
| | | Diplomatic missions | Operation of diplomatic and consular missions stationed abroad or at offices of international organisations. |
| | | Administration of developing countries' foreign aid | Support to administration of developing countries' foreign aid (including triangular and south-south cooperation). |
| | | General personnel services | Administration and operation of the civil service including policies, procedures and regulations. |
| | | Other general public services | Maintenance and storage of government records and archives, operation of government-owned or occupied buildings, central motor vehicle pools, government-operated printing offices, centralised computer and data processing services, etc. |
| | | National monitoring and evaluation | Operation or support of institutions providing national monitoring and evaluation. |
| | | Meteorological services | Operation or support of institutions dealing with weather forecasting. |
| | | National standards development | Operation or support of institutions dealing with national standards development. (Use code 16062 for statistical capacity-building.) |
| | | Executive office | Administration, operation or support of executive office. Includes office of the chief executive at all levels of government (monarch, governor-general, president, prime minister, governor, mayor, etc.). |
| | | Government and civil society statistics and data | Collection, production, management and dissemination of statistics and data related to Government & Civil Society. Includes macroeconomic statistics, government finance, fiscal and public sector statistics, support to development of administrative data infrastructure, civil society surveys. |
| | 15111 | Public finance management (PFM) | Fiscal policy and planning; support to ministries of finance; strengthening financial and managerial accountability; public expenditure management; improving financial management systems; budget drafting; inter-governmental fiscal relations, public audit, public debt. (Use code 15114 for domestic revenue mobilisation and code 33120 for customs). |
| | | Budget planning | Operation of the budget office and planning as part of the budget process. |

| DAC 5 | CRS | DESCRIPTION | Clarifications / Additional notes on coverage |
|---|---|---|---|
| | | National audit | Operation of the accounting and audit services. |
| | | Debt and aid management | Management of public debt and foreign aid received (in the partner country). For reporting on debt reorganisation, use codes 600xx. |
| | 15112 | Decentralisation and support to subnational government | Decentralisation processes (including political, administrative and fiscal dimensions); intergovernmental relations and federalism; strengthening departments of regional and local government, regional and local authorities and their national associations. (Use specific sector codes for decentralisation of sector management and services.) |
| | | Local government finance | Financial transfers to local government; support to institutions managing such transfers. (Use specific sector codes for sector-related transfers.) |
| | | Other central transfers to institutions | Transfers to non sector-specific autonomous bodies or state-owned enterprises outside of local government finance; support to institutions managing such transfers. (Use specific sector codes for sector-related transfers.) |
| | | Local government administration | Decentralisation processes (including political, administrative and fiscal dimensions); intergovernmental relations and federalism; strengthening local authorities. |
| | 15113 | Anti-corruption organisations and institutions | Specialised organisations, institutions and frameworks for the prevention of and combat against corruption, bribery, money-laundering and other aspects of organised crime, with or without law enforcement powers, e.g. anti-corruption commissions and monitoring bodies, special investigation services, institutions and initiatives of integrity and ethics oversight, specialised NGOs, other civil society and citizens' organisations directly concerned with corruption. |
| | 15114 | Domestic revenue mobilisation | Support to domestic revenue mobilisation/tax policy, analysis and administration as well as non-tax public revenue, which includes work with ministries of finance, line ministries, revenue authorities or other local, regional or national public bodies. (Use code 16010 for social security and other social protection.) |
| | | Tax collection | Operation of the inland revenue authority. |
| | | Tax policy and administration support | |
| | | Other non-tax revenue mobilisation | Non-tax public revenue, which includes line ministries, revenue authorities or other local, regional or national public bodies. |
| | 15125 | Public Procurement | Support to public procurement, including to create and evaluate legal frameworks; advice in establishing strategic orientation of public procurement policies and reforms; advice in designing public procurement systems and processes; support to public procurement institutions (including electronic procurement) as well as structures or initiatives to assess public procurement systems; and development of professional capacity of public procurement bodies and staff. |
| | 15130 | Legal and judicial development | Support to institutions, systems and procedures of the justice sector, both formal and informal; support to ministries of justice, the interior and home affairs; judges and courts; legal drafting services; bar and lawyers associations; professional legal education; maintenance of law and order and public safety; border management; law enforcement agencies, police, prisons and their supervision; ombudsmen; alternative dispute resolution, arbitration and mediation; legal aid and counsel; traditional, indigenous and paralegal practices that fall outside the formal legal system. Measures that support the improvement of legal frameworks, constitutions, laws and regulations; legislative and constitutional drafting and review; legal reform; integration of formal and informal systems of law. Public legal |

| DAC 5 | CRS | DESCRIPTION | Clarifications / Additional notes on coverage |
|-------|-----|-------------|-----------------------------------------------|
| | | | education; dissemination of information on entitlements and remedies for injustice; awareness campaigns. (Use codes 152xx for activities that are primarily aimed at supporting security system reform or undertaken in connection with post-conflict and peace building activities. Use code 15190 for capacity building in border management related to migration.) |
| | | Justice, law and order policy, planning and administration | Judicial law and order sectors; policy development within ministries of justice or equivalents. |
| | | Police | Police affairs and services. |
| | | Fire and rescue services | Fire-prevention and fire-fighting affairs and services. |
| | | Judicial affairs | Civil and criminal law courts and the judicial system, including enforcement of fines and legal settlements imposed by the courts and operation of parole and probation systems. |
| | | Ombudsman | Independent service representing the interests of the public by investigating and addressing complaints of unfair treatment or maladministration. |
| | | Immigration | Immigration affairs and services, including alien registration, issuing work and travel documents to immigrants. |
| | | Prisons | |
| | 15142 | Macroeconomic policy | Support to macroeconomic stability, debt sustainability and structural reforms. Includes technical assistance for strategic formulation of policies, laws and regulation; capacity building to enhance public sector development; policy-based funding. For fiscal policy and domestic revenue mobilisation use codes 15111 and 15114. |
| | 15150 | Democratic participation and civil society | Support to the exercise of democracy and diverse forms of participation of citizens beyond elections (15151); direct democracy instruments such as referenda and citizens' initiatives; support to organisations to represent and advocate for their members, to monitor, engage and hold governments to account, and to help citizens learn to act in the public sphere; curricula and teaching for civic education at various levels. (This purpose code is restricted to activities targeting governance issues. When assistance to civil society is for non-governance purposes use other appropriate purpose codes.) |
| | 15151 | Elections | Electoral management bodies and processes, election observation, voters' education. (Use code 15230 when in the context of an international peacekeeping operation.) |
| | 15152 | Legislatures and political parties | Assistance to strengthen key functions of legislatures/ parliaments including subnational assemblies and councils (representation; oversight; legislation), such as improving the capacity of legislative bodies, improving legislatures' committees and administrative procedures,; research and information management systems; providing training programmes for legislators and support personnel. Assistance to political parties and strengthening of party systems. |
| | 15153 | Media and free flow of information | Activities that support free and uncensored flow of information on public issues; activities that increase the editorial and technical skills and the integrity of the print and broadcast media, e.g. training of journalists. (Use codes 22010-22040 for provision of equipment and capital assistance to media.) |
| | 15160 | Human rights | Measures to support specialised official human rights institutions and mechanisms at universal, regional, national and local levels in their statutory roles to promote and protect civil and political, economic, social and cultural rights as defined in international conventions and covenants; translation of international human rights commitments into |

national legislation; reporting and follow-up; human rights dialogue. Human rights defenders and human rights NGOs; human rights advocacy, activism, mobilisation; awareness raising and public human rights education. Human rights programming targeting specific groups, e.g. children, persons with disabilities, migrants, ethnic, religious, linguistic and sexual minorities, indigenous people and those suffering from caste discrimination, victims of trafficking, victims of torture. (Use code 15230 when in the context of a peacekeeping operation and code 15180 for ending violence against women and girls. Use code 15190 for human rights programming for refugees or migrants, including when they are victims of trafficking.Use code 16070 for Fundamental Principles and Rights at Work, i.e. Child Labour, Forced Labour, Non-discrimination in employment and occupation, Freedom of Association and Collective Bargaining.)

**15170 — Women's rights organisations and movements, and government institutions**

Support for feminist, women-led and women's rights organisations and movements, and institutions (governmental and non-govermental) at all levels to enhance their effectiveness, influence and substainability (activities and core-funding). These organisations exist to bring about transformative change for gender equality and/or the rights of women and girls in developing countries. Their activities include agenda-setting, advocacy, policy dialogue, capacity development, awareness raising and prevention, service provision, conflict-prevention and peacebuilding, research, organising, and alliance and network building

**15180 — Ending violence against women and girls**

Support to programmes designed to prevent and eliminate all forms of violence against women and girls/gender-based violence. This encompasses a broad range of forms of physical, sexual and psychological violence including but not limited to: intimate partner violence (domestic violence); sexual violence; female genital mutilation/cutting (FGM/C); child, early and forced marriage; acid throwing; honour killings; and trafficking of women and girls. Prevention activities may include efforts to empower women and girls; change attitudes, norms and behaviour; adopt and enact legal reforms; and strengthen implementation of laws and policies on ending violence against women and girls, including through strengthening institutional capacity. Interventions to respond to violence against women and girls/gender-based violence may include expanding access to services including legal assistance, psychosocial counselling and health care; training personnel to respond more effectively to the needs of survivors; and ensuring investigation, prosecution and punishment of perpetrators of violence.

**15190 — Facilitation of orderly, safe, regular and responsible migration and mobility**

Assistance to developing countries that facilitates the orderly, safe, regular and responsible migration and mobility of people. This includes:• Capacity building in migration and mobility policy, analysis, planning and management. This includes support to facilitate safe and regular migration and address irregular migration, engagement with diaspora and programmes enhancing the development impact of remittances and/or their use for developmental projects in developing countries.• Measures to improve migrant labour recruitment systems in developing countries.• Capacity building for strategy and policy development as well as legal and judicial development (including border management) in developing countries. This includes support to address and reduce vulnerabilities in migration, and strengthen the transnational response to smuggling of migrants and preventing and combating trafficking in human beings.• Support to effective strategies to ensure international protection and the right to asylum.• Support to effective strategies to ensure access to justice and assistance for displaced persons.• Assistance to migrants for their safe, dignified, informed and voluntary return to their country of origin (covers only returns from another developing country; assistance to forced returns is excluded from ODA).• Assistance to migrants for their sustainable reintegration in their country of origin (use code 93010 for pre-departure assistance provided in donor countries in the context of voluntary returns). Activities that pursue first and foremost providers' interest are excluded from ODA. Activities addressing the root causes of forced displacement and irregular migration should not be coded here, but under their relevant sector of intervention. In addition, use code 15136 for support to countries' authorities for immigration affairs and services (optional), code 24050 for programmes aiming at reducing the sending costs of remittances, code 72010 for humanitarian aspects of assistance to refugees and internally displaced persons (IDPs) such as delivery of emergency services and humanitarian protection. Use code 93010 when expenditure is for the temporary sustenance of refugees in the donor country, including for their voluntary return and for their reintegration when support is provided in a donor country in connection with the return from that donor country (i.e. pre-departure assistance), or voluntary resettlement in a third developed country.

| DAC 5 | CRS | DESCRIPTION | Clarifications / Additional notes on coverage |
|---|---|---|---|
| 152 | | Conflict, Peace & Security | N.B. Further notes on ODA eligibility (and exclusions) of conflict, peace and security related activities are given in paragraphs 76-81 of the Directives. |
| | 15210 | Security system management and reform | Technical co-operation provided to parliament, government ministries, law enforcement agencies and the judiciary to assist review and reform of the security system to improve democratic governance and civilian control; technical co-operation provided to government to improve civilian oversight and democratic control of budgeting, management, accountability and auditing of security expenditure, including military budgets, as part of a public expenditure management programme; assistance to civil society to enhance its competence and capacity to scrutinise the security system so that it is managed in accordance with democratic norms and principles of accountability, transparency and good governance. [Other than in the context of an international peacekeeping operation (15230)]. |
| | 15220 | Civilian peace-building, conflict prevention and resolution | Support for civilian activities related to peace building, conflict prevention and resolution, including capacity building, monitoring, dialogue and information exchange. Bilateral participation in international civilian peace missions such as those conducted by the UN Department of Political Affairs (UNDPA) or the European Union (European Security and Defence Policy), and contributions to civilian peace funds or commissions (e.g. Peacebuilding Commission, Peacebuilding thematic window of the MDG achievement fund etc.). The contributions can take the form of financing or provision of equipment or civilian or military personnel (e.g. for training civilians).(Use code 15230 for bilateral participation in international peacekeeping operations). |
| | 15230 | Participation in international peacekeeping operations | Bilateral participation in peacekeeping operations mandated or authorised by the United Nations (UN) through Security Council resolutions, and conducted by international organisations, e.g. UN, NATO, the European Union (Security and Defence Policy security-related operations), or regional groupings of developing countries. Direct contributions to the UN Department for Peacekeeping Operations (UNDPKO) budget are excluded from bilateral ODA (they are reportable in part as multilateral ODA, see Annex 9). The activities that can be reported as bilateral ODA under this code are limited to: human rights and election monitoring; reintegration of demobilised soldiers; rehabilitation of basic national infrastructure; monitoring or retraining of civil administrators and police forces; security sector reform and other rule of law-related activities; training in customs and border control procedures; advice or training in fiscal or macroeconomic stabilisation policy; repatriation and demobilisation of armed factions, and disposal of their weapons; explosive mine |

| DAC 5 | CRS | DESCRIPTION | Clarifications / Additional notes on coverage |
| --- | --- | --- | --- |
| | | | removal. The enforcement aspects of international peacekeeping operations are not reportable as ODA. ODA-eligible bilateral participation in peacekeeping operations can take the form of financing or provision of equipment or military or civilian personnel (e.g. police officers). The reportable cost is calculated as the excess over what the personnel and equipment would have cost to maintain had they not been assigned to take part in a peace operation. Costs for military contingents participating in UNDPKO peacekeeping operations are not reportable as ODA. International peacekeeping operations may include humanitarian-type activities (contributions to the form of equipment or personnel), as described in codes 7xxxx. These should be included under code 15230 if they are an integrated part of the activities above, otherwise they should be reported as humanitarian aid. NB: When using this code, indicate the name of the operation in the short description of the activity reported. |
| | 15240 | Reintegration and SALW control | Reintegration of demobilised military personnel into the economy; conversion of production facilities from military to civilian outputs; technical co-operation to control, prevent and/or reduce the proliferation of small arms and light weapons (SALW) – see para. 80 of the Directives for definition of SALW activities covered. [Other than in the context of an international peacekeeping operation (15230) or child soldiers (15261)]. |
| | 15250 | Removal of land mines and explosive remnants of war | All activities related to land mines and explosive remnants of war which have benefits to developing countries as their main objective, including removal of land mines and explosive remnants of war, and stockpile destruction for developmental purposes [other than in the context of an international peacekeeping operation (15230)]; risk education and awareness raising; rehabilitation, reintegration and assistance to victims, and research and development on demining and clearance. Only activities for civilian purposes are ODA-eligible. |
| | 15261 | Child soldiers (prevention and demobilisation) | Technical co-operation provided to government – and assistance to civil society organisations – to support and apply legislation designed to prevent the recruitment of child soldiers, and to demobilise, disarm, reintegrate, repatriate and resettle (DDR) child soldiers. |
| 160 | | Other Social Infrastructure & Services | |
| | 16010 | Social Protection | Social protection or social security strategies, legislation and administration; institution capacity building and advice; social security and other social schemes; support programmes, cash benefits, pensions and special programmes for older persons, orphans, persons with disabilities, children, mothers with newborns, those living in poverty, without jobs and other vulnerable groups; social dimensions of structural adjustment. |
| | | Social protection and welfare services policy, planning and administration | Administration of overall social protection policies, plans, programmes and budgets including legislation, standards and statistics on social protection. |
| | | Social security (excl pensions) | Social protection shemes in the form of cash or in-kind benefits to people unable to work due to sickness or injury. |
| | | General pensions | Social protection schemes in the form of cash or in-kind benefits, including pensions, against the risks linked to old age. |
| | | Civil service pensions | Pension schemes for government personnel. |
| | | Social services (incl youth development and women+ children) | Social protection schemes in the form of cash or in-kind benefits to households with dependent children, including parental leave benefits. |
| | 16020 | Employment creation | Employment policy and planning; institution capacity building and advice; employment creation and income generation programmes; |

| DAC 5 | CRS | DESCRIPTION | Clarifications / Additional notes on coverage |
|-------|-----|-------------|-----------------------------------------------|
| | | | including activities specifically designed for the needs of vulnerable groups. |
| | 16030 | Housing policy and administrative management | Housing sector policy, planning and programmes; excluding low-cost housing and slum clearance (16040). |
| | 16040 | Low-cost housing | Including slum clearance. |
| | 16050 | Multisector aid for basic social services | Basic social services are defined to include basic education, basic health, basic nutrition, population/reproductive health and basic drinking water supply and basic sanitation. |
| | 16061 | Culture and recreation | Including libraries and museums. |
| | | Recreation and sport | |
| | | Culture | |
| | 16062 | Statistical capacity building | All statistical activities, such as data collection, processing, dissemination and analysis; support to development and management of official statistics including demographic, social, economic, environmental and multi-sectoral statistics; statistical quality frameworks; development of human and technological resources for statistics, investments in data innovation. Activities related to data and statistics in the sectors 120, 130 or 150 should preferably be coded under the voluntary purpose codes 12196, 13096 and 15196. Activities with the sole purpose of monitoring development co-operation activities, including if performed by third parties, should be coded under 91010 (Administrative costs). |
| | 16063 | Narcotics control | In-country and customs controls including training of the police; educational programmes and awareness campaigns to restrict narcotics traffic and in-country distribution. ODA recording of narcotics control expenditures is limited to activities that focus on economic development and welfare including alternative development programmes and crop substitution (see 31165 and 43050). Activities by the donor country to interdict drug supplies destroy crops or train or finance military personnel in anti-narcotics activities are not reportable. |
| | 16064 | Social mitigation of HIV/AIDS | Special programmes to address the consequences of HIV/AIDS, e.g. social, legal and economic assistance to people living with HIV/AIDS including food security and employment; support to vulnerable groups and children orphaned by HIV/AIDS; human rights of HIV/AIDS affected people. |
| | 16070 | Labour rights | Advocacy for international labour standards, labour law, fundamental principles and rights at work (child labour, forced labour, non-discrimination in the workplace, freedom of association and collective bargaining); formalisation of informal work, occupational safety and health. |
| | 16080 | Social dialogue | Capacity building and advice in support of social dialogue; support to social dialogue institutions, bodies and mechanisms; capacity building of workers' and employers' organisations. |
| 210 | | Transport & Storage | |
| | 21010 | Transport policy and administrative management | Transport sector policy, planning and programmes; aid to transport ministries; institution capacity building and advice; unspecified transport; activities that combine road, rail, water and/or air transport. Includes prevention of road accidents. Whenever possible, report transport of goods under the sector of the good being transported. |
| | | Transport policy, planning and administration | Administration of affairs and services concerning transport systems. |
| | | Public transport services | Administration of affairs and services concerning public transport. |

| DAC 5 | CRS | DESCRIPTION | Clarifications / Additional notes on coverage |
|---|---|---|---|
| | | Transport regulation | Supervision and regulation of users, operations, construction and maintenance of transport systems (registration, licensing, inspection of equipment, operator skills and training; safety standards, franchises, tariffs, levels of service, etc.). |
| | 21020 | Road transport | Road infrastructure, road vehicles; passenger road transport, motor passenger cars. |
| | | Feeder road construction | Construction or operation of feeder road transport systems and facilities. |
| | | Feeder road maintenance | Maintenance of feeder road transport systems and facilities. |
| | | National road construction | Construction or operation of national road transport systems and facilities. |
| | | National road maintenance | Maintenance of national road transport systems and facilities. |
| | 21030 | Rail transport | Rail infrastructure, rail equipment, locomotives, other rolling stock; including light rail (tram) and underground systems. |
| | 21040 | Water transport | Harbours and docks, harbour guidance systems, ships and boats; river and other inland water transport, inland barges and vessels. |
| | 21050 | Air transport | Airports, airport guidance systems, aeroplanes, aeroplane maintenance equipment. |
| | 21061 | Storage | Whether or not related to transportation. Whenever possible, report storage projects under the sector of the resource being stored. |
| | 21081 | Education and training in transport and storage | |
| 220 | | Communications | |
| | 22010 | Communications policy and administrative management | Communications sector policy, planning and programmes; institution capacity building and advice; including postal services development; unspecified communications activities. |
| | | Communications policy, planning and administration | |
| | | Postal services | Development and operation of postal services. |
| | | Information services | Provision of information services. |
| | 22020 | Telecommunications | Telephone networks, telecommunication satellites, earth stations. |
| | 22030 | Radio/television/print media | Radio and TV links, equipment; newspapers; printing and publishing. |
| | 22040 | Information and communication technology (ICT) | Computer hardware and software; internet access; IT training. When sector cannot be specified. |
| 230 | | Energy | |
| 231 | | Energy Policy | |
| | 23110 | Energy policy and administrative management | Energy sector policy, planning; aid to energy ministries and other governmental or nongovernmental institutions for activities related to the SDG7; institution capacity building and advice; tariffs, market building, unspecified energy activities; energy activities for which a more specific code cannot be assigned. |
| | | Energy sector policy, planning and administration | |
| | | Energy regulation | Regulation of the energy sector, including wholesale and retail electricity provision. |
| | 23181 | Energy education/training | All levels of training not included elsewhere. |
| | 23182 | Energy research | Including general inventories, surveys. |

| DAC 5 | CRS | DESCRIPTION | Clarifications / Additional notes on coverage |
|-------|-----|-------------|-----------------------------------------------|
| | 23183 | Energy conservation and demand-side efficiency | Support for energy demand reduction, e.g. building and industry upgrades, smart grids, metering and tariffs. For clean cooking appliances use code 32174. |
| 232 | | Energy generation, renewable sources | |
| | 23210 | Energy generation, renewable sources - multiple technologies | Renewable energy generation programmes that cannot be attributed to one single technology (codes 23220 through 23280 below). Fuelwood/charcoal production should be included under forestry 31261. |
| | 23220 | Hydro-electric power plants | Including energy generating river barges. |
| | 23230 | Solar energy for centralised grids | Including photo-voltaic cells, concentrated solar power systems connected to the main grid and net-metered decentralised solutions. |
| | 23231 | Solar energy for isolated grids and standalone systems | Solar power generation for isolated mini-grids, solar home systems (including integrated wiring and related appliances), solar lanterns distribution and commercialisation. This code refers to the power generation component only. |
| | 23232 | Solar energy - thermal applications | Solar solutions for indoor space and water heating (except for solar cook stoves 32174). |
| | 23240 | Wind energy | Wind energy for water lifting and electric power generation. |
| | 23250 | Marine energy | Including ocean thermal energy conversion, tidal and wave power. |
| | 23260 | Geothermal energy | Use of geothermal energy for generating electric power or directly as heat for agriculture, etc. |
| | 23270 | Biofuel-fired power plants | Use of solids and liquids produced from biomass for direct power generation. Also includes biogases from anaerobic fermentation (e.g. landfill gas, sewage sludge gas, fermentation of energy crops and manure) and thermal processes (also known as syngas); waste-fired power plants making use of biodegradable municipal waste (household waste and waste from companies and public services that resembles household waste, collected at installations specifically designed for their disposal with recovery of combustible liquids, gases or heat). See code 23360 for non-renewable waste-fired power plants. |
| 233 | | Energy generation, non-renewable sources | |
| | 23310 | Energy generation, non-renewable sources, unspecified | Thermal power plants including when energy source cannot be determined; combined gas-coal power plants. |
| | 23320 | Coal-fired electric power plants | Thermal electric power plants that use coal as the energy source. |
| | 23330 | Oil-fired electric power plants | Thermal electric power plants that use fuel oil or diesel fuel as the energy source. |
| | 23340 | Natural gas-fired electric power plants | Electric power plants that are fuelled by natural gas; related feed-in infrastructure (LNG terminals, gasifiers, pipelines to feed the plant). |
| | 23350 | Fossil fuel electric power plants with carbon capture and storage (CCS) | Fossil fuel electric power plants employing technologies to capture carbon dioxide emissions. CCS not related to power plants should be included under 41020. CCS activities are not reportable as ODA. |
| | 23360 | Non-renewable waste-fired electric power plants | Electric power plants that use non-biodegradable industrial and municipal waste as the energy source. |
| 234 | | Hybrid energy plants | |
| | 23410 | Hybrid energy electric power plants | Electric power plants that make use of both non-renewable and renewable energy sources. |
| 235 | | Nuclear energy plants | |

| DAC 5 | CRS | DESCRIPTION | Clarifications / Additional notes on coverage |
|---|---|---|---|
| | 23510 | Nuclear energy electric power plants and nuclear safety | See note regarding ODA eligibility of nuclear energy. |
| 236 | | Energy distribution | |
| | 23610 | Heat plants | Power plants which are designed to produce heat only. |
| | 23620 | District heating and cooling | Distribution of heat generated in a centralised location, or delivery of chilled water, for residential and commercial heating or cooling purposes. |
| | 23630 | Electric power transmission and distribution (centralised grids) | Grid distribution from power source to end user; transmission lines. Also includes storage of energy to generate power (e.g. pumped hydro, batteries) and the extension of grid access, often to rural areas. |
| | 23631 | Electric power transmission and distribution (isolated mini-grids) | Includes village grids and other electricity distribution technologies to end users that are not connected to the main national grid. Also includes related electricity storage. This code refers to the network infrastructure only regardless of the power generation technologies. |
| | 23640 | Retail gas distribution | Includes urban infrastructure for the delivery of urban gas and LPG cylinder production, distribution and refill. Excludes gas distribution for purposes of electricity generation (23340) and pipelines (32262). |
| | 23641 | Retail distribution of liquid or solid fossil fuels | |
| | 23642 | Electric mobility infrastructures | Includes electricity or hydrogen recharging stations for private and public transport systems and related infrastructure (except for rail transport 21030). |
| 240 | | Banking & Financial Services | |
| | 24010 | Financial policy and administrative management | Finance sector policy, planning and programmes; institution capacity building and advice; financial markets and systems. |
| | 24020 | Monetary institutions | Central banks. |
| | 24030 | Formal sector financial intermediaries | All formal sector financial intermediaries; credit lines; insurance, leasing, venture capital, etc. (except when focused on only one sector). |
| | 24040 | Informal/semi-formal financial intermediaries | Micro credit, savings and credit co-operatives etc. |
| | 24050 | Remittance facilitation, promotion and optimisation | Includes programmes aiming at reducing the sending costs of remittances. |
| | 24081 | Education/training in banking and financial services | |
| 250 | | Business & Other Services | |
| | 25010 | Business policy and administration | Public sector policies and institution support to the business environment and investment climate, including business regulations, property rights, non-discrimination, investment promotion, competition policy, enterprises law, private-public partnerships. |
| | 25020 | Privatisation | When sector cannot be specified. Including general state enterprise restructuring or demonopolisation programmes; planning, programming, advice. |
| | 25030 | Business development services | Public and private provision of business development services, e.g. incubators, business strategies, commercial linkages programmes and matchmaking services. Includes support to private organisations representing businesses, e.g. business associations; chambers of commerce; producer associations; providers of know-how and other business development services. For financial services use CRS |

| DAC 5 | CRS | DESCRIPTION | Clarifications / Additional notes on coverage |
|---|---|---|---|
| | | | codes 24030 or 24040. For SME development and for support to companies in the industrial sector use codes 32130 through 32172. For support to companies in the agricultural sector use code 31120. |
| | 25040 | Responsible business conduct | Support to policy reform, implementation and enforcement of responsible business conduct (RBC) principles and standards as well as facilitation of responsible business practices by companies. Includes establishing and enforcing a legal and regulatory framework to protect stakeholder rights and the environment, rewarding best performers; exemplifying RBC in government economic activities, such as state-owned enterprises' operations or public procurement; support to the implementation of the OECD Guidelines for MNEs, including disclosure, human rights, employment and industrial relations, environment, combating bribery, consumer interests, science and technology, competition and taxation. |
| 310 | | Agriculture, Forestry, Fishing | |
| 311 | | Agriculture | |
| | 31110 | Agricultural policy and administrative management | Agricultural sector policy, planning and programmes; aid to agricultural ministries; institution capacity building and advice; unspecified agriculture. |
| | 31120 | Agricultural development | Integrated projects; farm development. |
| | 31130 | Agricultural land resources | Including soil degradation control; soil improvement; drainage of water logged areas; soil desalination; agricultural land surveys; land reclamation; erosion control, desertification control. |
| | 31140 | Agricultural water resources | Irrigation, reservoirs, hydraulic structures, ground water exploitation for agricultural use. |
| | 31150 | Agricultural inputs | Supply of seeds, fertilizers, agricultural machinery/equipment. |
| | 31161 | Food crop production | Including grains (wheat, rice, barley, maize, rye, oats, millet, sorghum); horticulture; vegetables; fruit and berries; other annual and perennial crops. [Use code 32161 for agro-industries.] |
| | 31162 | Industrial crops/export crops | Including sugar; coffee, cocoa, tea; oil seeds, nuts, kernels; fibre crops; tobacco; rubber. [Use code 32161 for agro-industries.] |
| | 31163 | Livestock | Animal husbandry; animal feed aid. |
| | 31164 | Agrarian reform | Including agricultural sector adjustment. |
| | 31165 | Agricultural alternative development | Projects to reduce illicit drug cultivation through other agricultural marketing and production opportunities (see code 43050 for non-agricultural alternative development). |
| | 31166 | Agricultural extension | Non-formal training in agriculture. |
| | 31181 | Agricultural education/training | |
| | 31182 | Agricultural research | Plant breeding, physiology, genetic resources, ecology, taxonomy, disease control, agricultural bio-technology; including livestock research (animal health, breeding and genetics, nutrition, physiology). |
| | 31191 | Agricultural services | Marketing policies & organisation; storage and transportation, creation of strategic reserves. |
| | 31192 | Plant and post-harvest protection and pest control | Including integrated plant protection, biological plant protection activities, supply and management of agrochemicals, supply of pesticides, plant protection policy and legislation. |
| | 31193 | Agricultural financial services | Financial intermediaries for the agricultural sector including credit schemes; crop insurance. |
| | 31194 | Agricultural co-operatives | Including farmers' organisations. |

| DAC 5 | CRS | DESCRIPTION | Clarifications / Additional notes on coverage |
|-------|-----|-------------|-----------------------------------------------|
| | 31195 | Livestock/veterinary services | Animal health and management, genetic resources, feed resources. |
| 312 | | Forestry | |
| | 31210 | Forestry policy and administrative management | Forestry sector policy, planning and programmes; institution capacity building and advice; forest surveys; unspecified forestry and agro-forestry activities. |
| | 31220 | Forestry development | Afforestation for industrial and rural consumption; exploitation and utilisation; erosion control, desertification control; integrated forestry projects. |
| | 31261 | Fuelwood/charcoal | Sustainable forestry development whose primary purpose is production of fuelwood and charcoal. Further transformation of biomass in biofuels is coded under 32173. |
| | 31281 | Forestry education/training | |
| | 31282 | Forestry research | Including artificial regeneration, genetic improvement, production methods, fertilizer, harvesting. |
| | 31291 | Forestry services | |
| 313 | | Fishing | |
| | 31310 | Fishing policy and administrative management | Fishing sector policy, planning and programmes; institution capacity building and advice; ocean and coastal fishing; marine and freshwater fish surveys and prospecting; fishing boats/equipment; unspecified fishing activities. |
| | 31320 | Fishery development | Exploitation and utilisation of fisheries; fish stock protection; aquaculture; integrated fishery projects. |
| | 31381 | Fishery education/training | |
| | 31382 | Fishery research | Pilot fish culture; marine/freshwater biological research. |
| | 31391 | Fishery services | Fishing harbours; fish markets; fishery transport and cold storage. |
| 320 | | Industry, Mining, Construction | |
| 321 | | Industry | |
| | 32110 | Industrial policy and administrative management | Industrial sector policy, planning and programmes; institution capacity building and advice; unspecified industrial activities; manufacturing of goods not specified below. |
| | 32120 | Industrial development | |
| | 32130 | Small and medium-sized enterprises (SME) development | Direct support to improve the productive capacity and business management of micro, small and medium-sized enterprises in the industrial sector, including accounting, auditing, advisory services, technological transfer and skill upgrading. For business policy and institutional support use code 25010. For business development services through business intermediary organisations (e.g. business associations; chambers of commerce; producer associations; incubators; providers of know-how and other business development services) use CRS code 250xx. For farm and agricultural development use code 31120. |
| | 32140 | Cottage industries and handicraft | |

Here is the OECD classification, purpose codes are classified and agglomerated into sector codes

## B.    Software and hardware information

Here are the followings languages used throughout this research:

- R and Python for basic data treatment;
- Python for all the machine learning process;
- Rmarkdown for redaction purposes (presentations and article redaction).

The python scripts has been executed with an AMD EPYC 7302 16-core Processor 2,99 GHz (2 CPUs) and with 256 Gb of RAM.